# A Frequency-Domain Beamforming Procedure for Extracting Rayleigh Wave Attenuation Coefficients and Small-Strain Damping Ratio from 2D Ambient Noise Array Measurements

Aser Abbas [a, *], Mauro Aimar [b], Brady R. Cox [a], Sebastiano Foti [b]

[a] Utah State University, Department of Civil and Environmental Engineering, Logan, UT, USA, 84322.
[b] Politecnico di Torino, Department of Structural, Building and Geotechnical Engineering (DISEG), Torino, Italy.

**Abstract**

The small-strain damping ratio plays a crucial role in assessing the response of soil deposits to earthquake-induced ground motions and general dynamic loading. The damping ratio can theoretically be inverted for after extracting frequency-dependent Rayleigh wave attenuation coefficients from wavefields collected during surface wave testing. However, determining reliable estimates of in-situ attenuation coefficients is much more challenging than achieving robust phase velocity dispersion data, which are commonly measured using both active-source and ambient-wavefield surface wave methods. This paper introduces a new methodology for estimating frequency-dependent attenuation coefficients through the analysis of ambient noise wavefield data recorded by two-dimensional (2D) arrays of surface seismic sensors for the subsequent evaluation of the small-strain damping ratio. The approach relies on the application of an attenuation-specific wavefield conversion and frequency-domain beamforming. Numerical simulations are employed to verify the proposed approach and inform best practices for its application. Finally, the practical efficacy of the proposed approach is showcased through its application to field data collected at a deep, soft soil site in Logan, Utah, USA, where phase velocity and attenuation coefficients are extracted from surface wave data and then simultaneously inverted to develop deep shear wave velocity and damping ratio profiles.

*Corresponding author.
E-mail: aser.abbas@usu.edu (A. Abbas)

**Introduction**

The small-strain shear modulus ($G_{max}$) and small-strain damping ratio ($D$) form the starting point for many soil constitutive models and play a crucial role in assessing the response of soil deposits to earthquake-induced ground motions and general dynamic loading. $G_{max}$ is directly related to the in-situ shear wave velocity ($V_s$), and it represents the soil stiffness and its resistance to deformation under applied shear stress. $D$ characterizes the energy dissipation properties of the material. The influence of $D$ on the amplitude and frequency content of seismic waves has been recognized since at least 1940 (Ricker, 1940), with subsequent research establishing it as a pivotal parameter for seismic site response studies and for modeling ground-borne vibrations (e.g., Anderson et al., 1996; Tao and Rathje, 2019; Papadopoulos et al., 2019; Foti et al., 2021). Despite its significance, the in-situ estimation of $D$ has received far less attention when compared to measurements of $V_s$ (Parolai, 2014). $D$ can theoretically be inverted for after extracting frequency-dependent Rayleigh wave phase velocity and attenuation coefficients (α) from wavefields collected during surface wave testing (Lai, 1998; Foti, 2004). However, in-situ α values are generally much more difficult to reliably measure than phase velocities (Haendel et al., 2016; Parolai et al., 2022), which are commonly measured using both active-source and ambient-wavefield surface wave methods.

This paper introduces a new noninvasive method to estimate frequency-dependent Rayleigh wave α using ambient noise wavefield data collected with two-dimensional (2D) arrays of surface seismic sensors for the subsequent evaluation of $D$. The approach relies on frequency-domain beamforming (FDBF) and applies an attenuation-specific wavefield conversion, known as the FDBFa approach. While Aimar et al. (2024a) previously used this approach for active-source surface wave testing, it has not been applied to ambient noise surface wave testing. In this paper, we introduce a new method called the noise FDBFa (NFDBFa) approach and document its development and application.

The subsequent sections of this paper are organized as follows: first, we cover important background information on attenuation and damping. Second, we present a concise overview of the FDBF technique introduced by Lacoss et al. (1969) and the FDBFa wavefield conversion methodology proposed by Aimar et al. (2024a), along with the integration of these methods within our proposed NFDBFa approach. Then, synthetic studies are presented to showcase the capabilities of the proposed NFDBFa approach and inform best practices for its application. The synthetic studies offer valuable insights into the influence of 2D array size and proximity to noise sources on attenuation estimates. For example, it is demonstrated that the optimal 2D ambient noise array design principles for attenuation estimation differ from the principles governing 2D array design for phase velocity estimation. Finally, we demonstrate the practical utility of our proposed NFDBFa technique through a field application at a deep, soft soil site in Logan, Utah, USA. In this field application, phase velocity and attenuation coefficients are extracted from surface wave data and then simultaneously inverted to develop deep $V_s$ and $D$ profiles. The good agreement observed between the attenuation estimates derived from our new NFDBFa technique and those obtained through the standard FDBFa analysis of active-source data collected using the multichannel analysis of surface waves (MASW) provides compelling evidence of the effectiveness of our new ambient noise approach.

**Background**

The attenuation of seismic waves in a continuum is related to the damping ratios of both compression waves ($D_p$) and shear waves ($D_s$). Surprisingly, little is known about the relative relationship between $D_p$ and $D_s$, and one can find instances in the literature where researchers have assumed $D_p = D_s$ (Badsar et al., 2010; Verachtert et al., 2017; Aimar et al., 2024b), $D_p > D_s$ (Bergamo et al., 2023), and $D_p < D_s$ (Xia et al., 2002). In this paper, when '$D$' is used without a subscript, it implies that the statement or equation is valid for both $D_p$ and $D_s$. The damping ratio ($D$) is commonly used in engineering, while its inverse, the quality factor ($Q$), where $Q^{-1} = 2D$, is more prevalent in seismological and geophysical literature (Foti, 2004). Consequently, $Q$, being the inverse of $D$, also varies for compressional waves ($Q_p$) and shear waves ($Q_s$).

Seismic wave attenuation is commonly attributed to three mechanisms: material damping, geometric spreading, and apparent attenuation (Zywicki, 1999). Material damping, or anelastic attenuation, arises from the collective interaction of diverse factors (Johnston et al., 1979). These factors encompass frictional losses among solid particles and fluid flow losses due to the relative motion between solid and fluid phases, a phenomenon particularly notable in coarse-grained soils (Biot, 1956; Walsh, 1966 and 1968; Stoll, 1974). Fine-grained soils, however, showcase more intricate phenomena influenced by electromagnetic interactions between water dipoles and microscopic solid particles (Lai, 1998). This intrinsic material damping is typically approximated as frequency-independent, especially within the seismic frequency band, primarily spanning 0.1 to 10 Hz (Aki and Richards, 1980; Shibuya et al., 1995). Material damping gives rise to a cyclic stress-strain curve exhibiting a hysteretic loop and is commonly referred to as hysteretic damping (Rix et al., 2000; Parolai et al., 2022).

Geometric or radiation damping involves the spread of a fixed amount of energy over a broader area or volume as the wavefront moves away from the source. Take, for instance, a harmonic unit point load applied along the normal direction to the surface of a homogeneous and isotropic half-space; this perturbation generates both body waves and Rayleigh waves. The body waves propagate radially from the source, forming a hemispherical wave front, while Rayleigh waves travel outward along a cylindrical wave front. As these waves travel, they traverse an expanding volume of material, leading to a decrease in energy density as the distance from the source increases. The amplitude of the body waves attenuates in proportion to the ratio of $r^{-1}$ (where $r$ is the radial distance from the source), except when along the surface of the half-space. In that case, the amplitude attenuates proportionally to $r^{-2}$. Conversely, the amplitude of the Rayleigh waves attenuates as $r^{-0.5}$ (Lamb, 1904; Ewing et al., 1957; Richart et al., 1970). Consequently, at substantial distances from the surface source, the dominant influence on overall particle motion stems from the surface wavefield (Lai, 1998). It is worth mentioning that these geometric spreading rules do not hold with transient waveforms (Keilis-Borok, 1989) or non-homogeneous media (Lai, 1998).

Apparent attenuation includes wave scattering, which arises from the interaction of waves with heterogeneities along the seismic path (O'Doherty and Anstey, 1971; Spencer et al., 1977), as well as the reflection and transmission of seismic waves at interfaces and mode conversions (Rix et al., 2000). Therefore, apparent attenuation is highly site-specific and difficult to generalize.

Both laboratory tests and in-situ methods have been proposed to estimate $D$. Laboratory tests, such as the resonant column (ASTM D4015-21), are valuable for parametrically studying the material/intrinsic damping ratio, but they cannot capture the other two mechanisms contributing to the attenuation of seismic waves in situ. Conversely, the damping ratio estimates obtained using in-situ methods are influenced by all the seismic wave damping mechanisms mentioned above (Parolai et al., 2022). In-situ methods also have the advantage of assessing soil characteristics in their natural and undisturbed state (Rix et al., 2000). Additionally, in-situ tests encompass a greater soil volume, effectively reducing result biases that might arise from localized variations in soil properties (Badsar et al., 2010). Furthermore, they provide parameter estimates on a spatial scale relevant to common engineering applications (e.g., Comina et al. 2011). In the scope of estimating $D$, in-situ methods can be dissected into two categories: invasive and noninvasive methods. Invasive methods encompass techniques such as cross-hole testing (Jongmans, 1989; Hall and Bodare, 2000) and downhole testing (Michaels, 1998; Crow et al., 2011). Noninvasive methods, particularly surface wave techniques, offer numerous advantages. By situating sensors at the ground surface, surface wave methods accelerate data acquisition, minimize costs, streamline validation of soil-receiver coupling, and encompass a frequency range closely aligned with those pertinent to earthquake engineering applications (Rix et al., 2000; Verachtert et al., 2017; and Parolai et al., 2022).

Surface wave testing became popular in the 1980's as an effective way to non-invasively develop 1D layering and $V_s$ profiles for both soil deposits and pavement systems (e.g., Nazarian et al., 1983; Stokoe et al., 1989). Typically, the use of surface wave methods involves acquiring experimental phase velocity dispersion data through active-source methods, ambient-noise methods, or a combination of both (Tokimatsu, 1995). These dispersion data are then inverted to obtain layered subsurface models, with the primary goal of resolving changes in $V_s$. The combined use of active-source and ambient-noise methods facilitates the generation of dispersion data across a wide frequency range, which enables resolution of both near-surface and deeper layers. Active sources predominantly produce energy concentrated at higher frequencies, typically ranging from several Hertz to perhaps 100 Hertz, with limited energy generation below 5-10 Hz for small sources like sledgehammers and drop weights. Consequently, the effective profiling depth using active-source methods is often constrained to approximately 15 to 40 m, contingent on the subsurface velocity and source mass (Foti et al., 2018). The primary hindrance to achieving increased penetration depths lies in generating lower frequency (i.e., longer wavelength) waves with affordable and highly-portable sources. This difficulty is circumvented by ambient-noise methods, which do not involve the active generation of wave energy. Instead, they rely on ground motions induced by cultural noise and microtremors (i.e., ambient noise), encompassing an abundance of low-frequency components (Lai, 1998). Consequently, ambient-noise surveys offer valuable insights for deep characterization, extending to depths of hundreds of meters or more (Foti et al., 2014; Teague et al., 2018a). Nevertheless, the spectral power of microtremors is generally low at higher frequencies (Peterson, 1993), which limits their ability to resolve changes in stiffness near the ground surface (Tokimatsu, 1995; Foti et al., 2014). Combining both active and ambient-noise measurements offers a solution to overcome this limitation.

Ambient-noise surveys typically employ 2D arrays of surface seismic sensors due to the a-priori unknown location of the ambient noise sources. Unlike linear arrays, 2D arrays allow for the determination of wave propagation direction, which is necessary for resolving the true phase velocity (Cox and Beekman, 2011). While 2D ambient noise array measurements have been referred to using several names, in this paper we will refer to them as microtremor array measurements (MAM; Ohrnberger et al., 2004; Teague et al., 2018b). A schematic representation of a typical survey utilizing both active and ambient-noise arrays is presented in Figure 1a. The active-source array in Figure 1a is in accordance with the MASW method (Park et al., 1999), utilizing a linear array of receivers to capture the wavefield generated by active sources off each end of the array. Example waveforms recorded by 24 receivers placed in-line with one of the active sources to the left of the array are depicted in Figure 1b. The ambient wavefield array depicted in Figure 1a is in accordance with MAM testing, where surface sensors are deployed in a 2D circular pattern (note that other 2D geometries are also permissible). Example ambient noise waveforms recorded by nine sensors in the circular array are depicted in Figure 1d. Figure 1c schematically illustrates phase velocity dispersion data that are commonly extracted from active-source MASW waveforms and ambient noise MAM waveforms using various well-known wavefield transformation techniques (Vantassel and Cox, 2022). Examples of these techniques include frequency-domain beamforming (FDBF; Lacoss et al., 1969), high-resolution frequency-wavenumber (f-k) spectrum analysis (Capon 1969), cylindrical FDBF (Zywicki 1999; Zywicki and Rix, 2005), and Rayleigh three-component beamforming (Wathelet et al., 2018). The combined dispersion data from MASW and MAM spans a wide frequency range, encompassing both low frequencies obtained from the MAM testing and high frequencies obtained from the MASW testing, with some overlap in between. The phase velocity dispersion data are then typically used to solve the parameter identification problem (i.e., inversion) and obtain 1D $V_s$ profiles of the subsurface (Foti et al., 2018; Vantassel and Cox, 2021). Note that the inversion step and resulting $V_s$ profiles are not illustrated schematically in Figure 1.

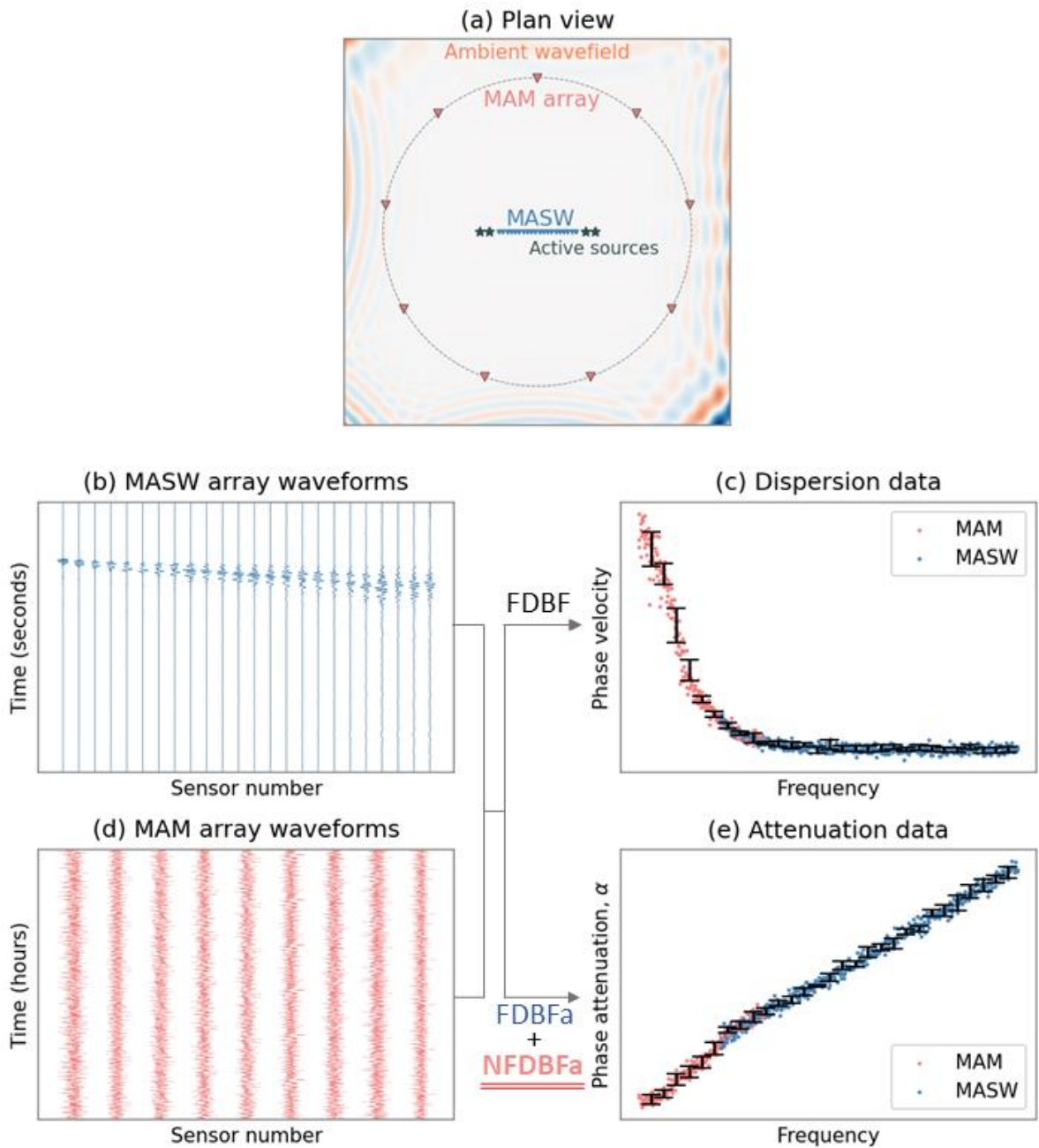

**Figure 1.** Schematic illustrating the data acquisition and processing stages of active-source and ambient-wavefield surface wave testing used to extract phase velocity and phase attenuation data. Panel (a) presents a typical acquisition setup consisting of concentric MASW and MAM arrays, featuring active sources for the MASW array and an ambient wavefield for the MAM array. Panel (b) shows waveforms from a single active-source location collected using the MASW array, while Panel (c) presents the combined phase velocity dispersion data resulting from MASW and MAM Frequency Domain Beamforming (FDBF) processing. Panel (d) depicts the ambient noise waveforms collected from the MAM array. In Panel (e), phase attenuation data processed through active-source FDBFa and ambient-wavefield NFDBFa techniques are illustrated.

As noted above, much more effort has been devoted to extracting phase velocity information from surface wave approaches than to extracting attenuation information. Nonetheless, multiple active-source methods have been developed to estimate the attenuation of surface waves. The methods introduced by Lai (1998), Lai et al. (2002), Rix et al. (2000), Xia et al. (2002), and Foti (2004) are founded on assessing the spatial decay of Rayleigh waves, a phenomenon that is influenced by both $D_p$ and $D_s$, as described by Aki and Richards (1980). These approaches assume the dominance of a single Rayleigh wave mode of propagation. Consequently, they might yield inaccurate results in soil profiles where multiple surface wave modes significantly contribute to the wavefield propagation (Rix et al. 2001). Badsar et al. (2010) introduced the half-power bandwidth method, originally developed in the field of mechanical and structural dynamics to determine the modal damping ratio of a structure, to assess Rayleigh modal attenuation by analyzing the width of the Rayleigh peaks in the f-k domain. Verachtert et al. (2017) employed the circle fit method, originally developed to determine eigenfrequencies and modal damping ratios in structural dynamics (Ewins 1984), to estimate multimodal Rayleigh dispersion and attenuation curves. Both the half-power bandwidth and circle fit methods facilitated the determination of modal attenuation curves from multimode wavefields (Verachtert et al., 2017). Recently, Aimar et al. (2024a) introduced an innovative technique that combines a novel wavefield conversion approach coupled with FDBF (Lacoss et al., 1969) for processing active-source data collected using MASW to estimate the frequency-dependent $α$ values. They called this the FDBF attenuation (FDBFa) method. Notably, the wavefield conversion proposed by Aimar et al. (2024a) to extract $α$ differs from the conventional wavefield transformations commonly used to go from the time-distance domain to the f-k domain, as detailed in the following section. To avoid confusion, we will refer to the wavefield transformation proposed by Aimar et al. (2024a) as 'wavefield conversion,' while reserving the term 'wavefield transformation' specifically for the more common f-k domain transformations used to extract phase velocity data.

While important research on extracting phase attenuation coefficients using active-source methods is ongoing, similar to phase velocity data, combining active-source and ambient noise methods is desirable for resolving attenuation data over a broader frequency band. The majority of ambient noise techniques aimed at estimating the attenuation of surface waves were developed for regional-scale estimation (Haendel et al., 2016; Parolai et al., 2022). Only a limited number of approaches have considered local scales that hold relevance for engineering purposes, like site-specific seismic ground response analyses or dynamic vibration studies. These local-scale approaches are predominantly based on retrieving attenuation properties from the cross-correlation of seismic noise (e.g., Albarello and Baliva, 2009; Parolai, 2014; Haendel et al., 2016). Albarello and Baliva (2009) proposed a methodology that reconstructs the Green's function based on the temporal derivative of averaged cross-correlations from noise recordings obtained by pairs of geophones, thereby incorporating attenuation effects into the process. They further validated this approach by demonstrating its potential in estimating attenuation coefficients at two distinct sites. Parolai (2014) estimated the Rayleigh phase velocity and attenuation coefficients by fitting a damped zero-order Bessel function, introduced by Prieto et al. (2009), using data generated from the space correlation function introduced by Aki (1957). To mitigate the impact of uneven source distribution on cross-correlations, Haendel et al. (2016) employed a higher-order noise cross-correlation technique to extract the phase velocity and attenuation coefficient of Love waves. They illustrated that their approach yields correlation functions with higher signal-to-noise ratios compared to simple noise cross-correlations.

The importance of seismic noise cross-correlation methods cannot be underestimated. Nonetheless, in theory, the reconstruction of the full Green's function requires the noise wavefield energy to be equally partitioned in all directions (Sánchez-Sesma and Campillo, 2006; Snieder et al., 2007). This is a highly specific condition that rarely met rigorously by ambient noise on Earth (Cupillard and Capdeville, 2010; Tsai, 2011; Haendel et al., 2016). Furthermore, while travel time measurements from cross-correlation of ambient noise are theoretically understood, amplitude measurements lack a corresponding theoretical background, except when the noise is equipartitioned (Snieder et al., 2007; Tsai, 2011). Studies by Cupillard and Capdeville (2010) and Tsai (2011) have shown that attenuation estimates using cross-correlations are significantly influenced by the distribution of the noise sources. In light of the challenges posed by the equipartitioning condition for the reconstruction of the full Green's function in ambient noise studies (Sánchez-Sesma and Campillo, 2006; Snieder et al., 2007), and considering the limitations highlighted by Cupillard and Capdeville (2010) and Tsai (2011) regarding the influence of noise source distribution on attenuation estimates, we introduce a paradigm-shifting approach herein for calculating attenuation coefficients from ambient noise. This novel method not only eliminates the need for an equipartitioned noise wavefield, but also remains robust in the face of uneven noise source distribution, marking a departure from existing methodologies.

This paper builds upon Aimar et al.'s (2024a) work on developing an FDBFa technique for estimating $\alpha$ from active-source MASW testing and expands the FDBFa approach to ambient noise data recorded using MAM. Importantly, using an FDBF approach enables the actual direction of ambient noise propagation to be determined for each noise window and frequency, and does not require equipartitioning of ambient noise energy. Furthermore, using an FDBF approach enables the phase attenuation data generated from MASW and that from MAM to be combined in order to generate phase attenuation data spanning a broader frequency range, as illustrated schematically in Figure 1e. The experimental dispersion and attenuation data can then be combined and inverted to determine not only the $V_s$ profile but also the $D$ profile of the subsurface to greater depths. This inversion of dispersion and attenuation data to obtain $V_s$ and $D$ profiles can be carried out either sequentially, as demonstrated in the work of Rix et al. (2000), or simultaneously, as shown by both Lai (1998) and Aimar et al. (2024b).

**Wavefield conversion Proposed by Aimar et al., (2024a)**

The method introduced by Aimar et al. (2024a) to estimate Rayleigh wave attenuation ($\alpha$) assumes that the recorded wavefield is dominated by planar surface waves, specifically Rayleigh waves observed in the far field, with a dominant propagation mode. Several techniques have been developed to estimate the wavenumber ($k$) and therefore the phase velocity from such wavefields (e.g., Lacoss et al., 1969; Capon, 1969; Zywicki and Rix, 2005; Wathelet et al., 2018). Aimar et al. (2024a) harnessed this concept and introduced a novel wavefield conversion approach that provides a pathway for calculating $\alpha$ by utilizing methods from existing literature originally developed for estimating $k$. The methodology involves converting the recorded wavefield into a function interpreted as a pseudo-wave. This pseudo-wave exhibits dispersion characteristics reflecting the phase attenuation of the original wave. The determination of $\alpha$ then becomes straightforward through the application of existing techniques for estimating $k$.

Consider the harmonic, exponentially decaying displacement wavefield, $U(r)$, depicted in Figure 2a and expressed by Equation 1. This wavefield is observed at many discrete distances at a specific moment in time and is induced by the passage of a monochromatic plane wave. Within this wavefield, $\alpha$ governs the amplitude decay resulting from material damping in accordance with a viscoelastic constitutive model. When the wavefield is plotted as log amplitude versus radial distance ($r$) from the source, the slope of the amplitude decay is $\alpha$, as illustrated in Figure 2b. When the wavefield is plotted as phase angle versus *r*, *k* denotes the slope of the unwrapped phase (i.e., the linear phase shift), as shown in Figure 2c. Aimar et al. (2024a) proposed raising the recorded wavefield, *U(r)*, to the power of the imaginary number, i (see Equation 2). Consequently, a pseudo displacement wavefield, $v(r)$, is generated, wherein the wavenumber is modulated by *α*, signifying that when the unwrapped phase of the converted wavefield is graphed against radial distance, the slope of that phase corresponds to the value of *α* (refer to Figure 2d). Conversely, when the log amplitude of the converted wavefield is plotted against distance, the slope manifests as *k*, with an inverted sign (refer to Figure 2e). This wavefield conversion allows for estimating *α* using any of the already established and common wavefield transformation techniques for calculating *k* (e.g., f-k or FDBF methods).

$$U(r) = e^{-\alpha r} e^{-ikr} \qquad (1)$$

$$v(r) = [U(r)]^i = e^{-i\alpha r} e^{kr} \qquad (2)$$

This wavefield conversion can also be extended to a broadband wavefield, comprised of a superposition of monochromatic plane waves by exponentiating the wavefield in the frequency domain with the power of the imaginary number. To address numerical artifacts introduced by the wrapped phase on the pseudo wavefield, Aimar et al. (2024a) recommended normalizing *v(r)* by its amplitude on a frequency-by-frequency basis. Aimar et al. (2024a) showed that this wavefield conversion can be successfully applied to active-source wavefields recorded using MASW as a means to estimate *α*. In this paper, we extend this approach to estimate *α* from ambient noise wavefields recorded using MAM arrays, employing the FDBF technique introduced by Lacoss et al. (1969).

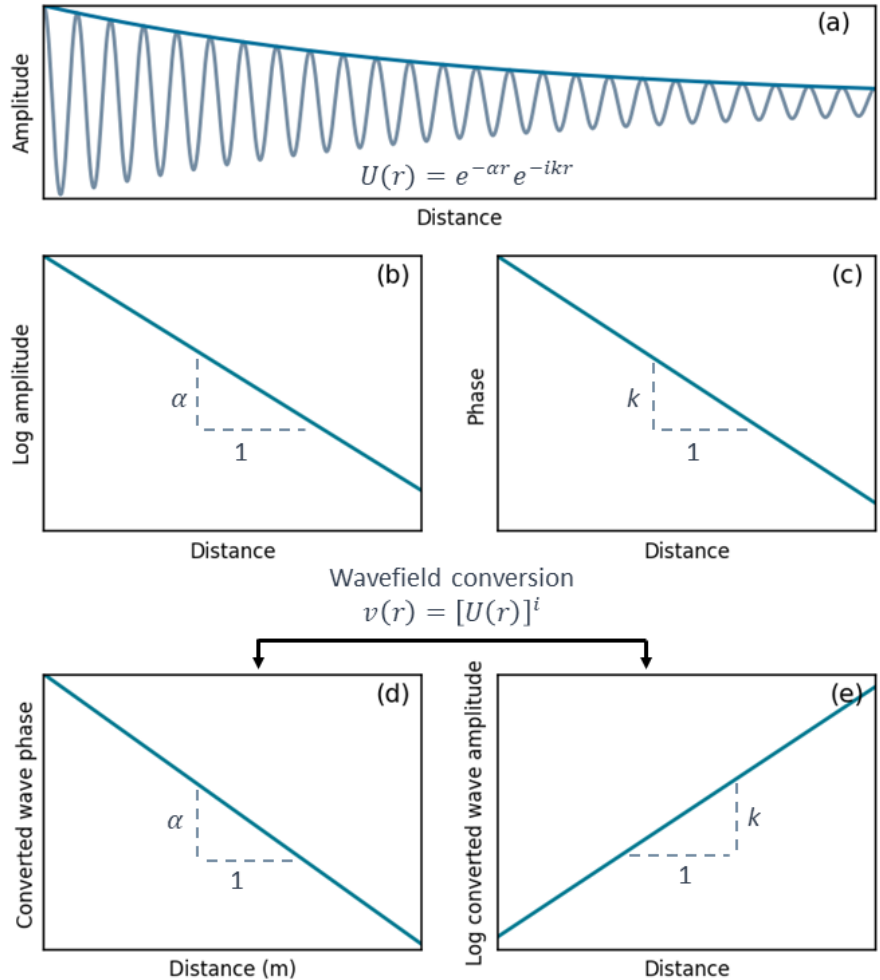

**Figure 2.** Schematic illustrating the wavefield conversion approach proposed by Aimar et al. (2024a) to extract attenuation coefficients ($\alpha$). Panel (a) displays the particle displacement of a monochromatic wave experiencing exponential amplitude decay with distance, indicative of material damping in a viscoelastic constitutive model. Panel (b) depicts linear amplitude decay in log amplitude versus linear distance space, where the slope represents the phase attenuation coefficient. In Panel (c), the modulation of the unwrapped phase slope with distance by the wavenumber ($k$) is demonstrated. Panel (d) illustrates the modulation of the unwrapped phase slope by the phase attenuation coefficient in the converted wavefield. Panel (e) showcases the control of the slope of the log amplitude decay with linear distance by the wavenumber, albeit with an inverted sign.

**Noise Frequency Domain Beam Forming - attenuation (NFDBFa)**

The inherent challenge in ambient noise measurements stems from the lack of *a priori* information about the source location or the direction of wave propagation, necessitating the use of spatial 2D arrays to determine the noise propagation directions during post-processing (Zywicki, 1999). As ambient noise wavefields operate in two spatial dimensions (e.g., x and y), it is necessary to represent the wavenumber using 2D vectors (Johnson and Dudgeon, 1993; Zywicki, 1999), where $\vec{k} = k_x \hat{\imath} + k_y \hat{\jmath}$, and $\hat{\imath}$ and $\hat{\jmath}$ are unit vectors in the x and y directions, respectively. Similarly, $\vec{\alpha}$ is also expressed as a 2D vector (i.e., $\vec{\alpha} = \alpha_x \hat{\imath} + \alpha_y \hat{\jmath}$) in this paper. Beamforming refers to a diverse set of array processing algorithms that concentrate the signal-capturing capabilities of the array in a specific direction. The fundamental concept behind beamforming is straightforward: when a propagating signal exists within an array's aperture, the outputs of the sensors, delayed by appropriate amounts and added together, enhance the coherent signal while mitigating the incoherent signal from waves propagating in different directions. The delays that enhance the signal are directly linked to the time it takes for the signal to travel between sensors (Johnson and Dudgeon, 1993). Delays in the time domain correspond to linear phase shifts in the frequency domain, providing information about the wavenumber. FDBF calculations are exclusively performed within the frequency domain. Applying FDBF to the original wavefield, $U(\vec{r})$, provides information about $\vec{k}$, which informs the estimation of the phase velocity. This paper aims to demonstrate that applying FDBF to the converted, normalized pseudo wavefield, $v(\vec{r})$, informs the estimation of $\vec{\alpha}$. Henceforth, in this paper, we will denote FDBF applied to the converted noise wavefield as NFDBFa, emphasizing its role in estimating the phase attenuation from ambient noise.

In the NFDBFa approach, the first step is to partition the noise data collected by a 2D array of *m* sensors into *B* time windows. The *m* sensors are located at the ground surface at coordinates $(x_i, y_i)$ denoted by the vector $\vec{r_i}$, where *i* varies from 1 to *m*. For each time window, Fourier spectra are calculated. Following this, the complex number at each frequency in the spectra is exponentiated to the imaginary power. Then, each exponentiated complex number is normalized by dividing it by its absolute amplitude. This process is conducted to obtain the normalized spectra of the pseudo wavefield (Aimar et al., 2024a). These spectra are then used to compute the Hermitian symmetric spatio-spectral correlation matrix, $R_{ij}$, with *i* and *j* representing indices of the *m* sensors in the 2D array, using Equation 3:

$$R_{ij}(\omega) = \frac{1}{B}\sum_{n=1}^{B} v_{i,n}(\omega)\, v_{j,n}^*(\omega) \qquad (3)$$

where $R_{ij}(\omega)$ is the averaged pseudo cross-power spectrum between the $i^{th}$ and $j^{th}$ sensors in the array across all windows, $v_{i,n}(\omega)$ is the normalized pseudo spectra of the $i^{th}$ sensor's data in the $n^{th}$ window, * indicates complex conjugation, and $\omega$ is the angular frequency. Despite being frequency-dependent, the spatio-spectral correlation matrix conveys spatial wavefield properties. Power within specific frequency-phase attenuation ($f$-$\vec{\alpha}$) pairs is determined by steering the array towards various directions and potential phase attenuation values. Array steering involves exponential phase shift vectors determined by trial $\vec{\alpha}$ values in pseudo space, as given by Equation 4:

$$e(\vec{\alpha}) = [\exp(-i\vec{\alpha}\cdot\vec{r_1}), \ldots, \exp(-i\vec{\alpha}\cdot\vec{r_m})]^T \tag{4}$$

where $e(\vec{\alpha})$ is a steering vector associated with a trial $\vec{\alpha}$ and $T$ denotes the transpose of the vector. The power in a particular $f$-$\vec{\alpha}$ pair, $P_{NFDBFa}(\vec{\alpha}, \omega)$, is estimated by multiplying $R_{ij}(\omega)$ by $e(\vec{\alpha})$ and summing the total power over all sensors, as given by Equation 5:

$$P_{NFDBFa}(\vec{\alpha}, \omega) = e^H(\vec{\alpha})R_{ij}(\omega)e(\vec{\alpha}) \tag{5}$$

where $H$ indicates the Hermitian transpose. The steering vectors aim to align the array with plane waves propagating from a specified direction and phase attenuation for each frequency. The successful alignment results in a peak within the $P_{NFDBFa}(\vec{\alpha}, \omega)$ pseudo-spectrum estimate. Thus, the NFDBFa technique presented herein allows for estimating $\alpha$ from ambient noise data without requiring an equipartitioned wavefield.

Even though there are similarities between the FDFBa method proposed by Aimar et al. (2024a) for estimating $\alpha$ using an MASW test setup and the NFDBFa method introduced in this study, there are notable differences between the two. Part of the difference is a consequence of the inherent dissimilarities between MASW and MAM. In the FDBFa method, the source location is predetermined and the array is aligned with the source, simplifying the problem and enabling the use of wavefield transformations like cylindrical frequency domain beamforming (Zywicki and Rix, 2005). Moreover, the signal-to-noise ratio can be readily enhanced by time-domain or frequency-domain stacking, as advocated by Vantassel and Cox (2022) and Foti et al., (2018). Additionally, dispersion and attenuation uncertainties can be quantified using the multiple source offset approach proposed by Cox and Wood (2011). In contrast, the NFDBFa approach developed in this study encounters distinct challenges, primarily arising from the *a priori* unknown location of the source(s). This necessitates the utilization of 2D arrays and involves azimuthally scanning the 2D space to ascertain the direction of the most coherent source of energy at each frequency for each window. Furthermore, the enhancement of the coherent noise-to-incoherent noise ratio involves averaging multiple time windows, while uncertainty quantification involves analyzing various time blocks, each composed of different windows. Thus, in the NFDBFa approach, data are recorded for significantly longer durations (i.e., hours) compared to FDBFa (i.e., seconds). Additionally, the NFDBFa approach relies on measurements of ambient noise, which is typically assumed to be generated by sources located in the far field. If this assumption holds true, it helps to mitigate the impact of geometric spreading, which plays a significant role on attenuation estimates near an active source (Badsar, 2012). Nearfield noise sources lead to complications in extracting accurate attenuation estimates, as discussed in greater detail below.

Figure 3 presents examples of the FDBF and NFDBFa responses obtained from a synthetic wavefield recorded by a ten-receiver circular MAM array for a single frequency and single time window. The array comprises nine sensors equally spaced on the perimeter of the circle and one sensor in the middle. The FDBF method is utilized to estimate $\vec{k}$ and the NFDBFa method is utilized to estimate $\vec{\alpha}$. In Figures 3a and 3b, the results of applying the FDBF technique to the original noise wavefield recorded by the array are depicted. Figure 3a illustrates the $f$-$\vec{k}$ spectrum at the considered frequency in a 2D wave number space ($k_x$-$k_y$). Stronger powers are represented by a darker purple color. This spectrum offers insights into the power and vector velocities of propagating waves. In this example, a wave propagates along the x-axis with a velocity represented by a vector wave number $\vec{k}$ at the chosen frequency. Consequently, a spectrum peak emerges on the positive $k_x$ axis at a distance of $|\vec{k}|$ from the origin. The associated phase velocity can be calculated as $V_r = 2\pi f/|\vec{k}|$, and the wavelength, $\lambda$, can be determined as $\lambda = 2\pi/|\vec{k}|$. Figure 3b illustrates the cross-section a-a from Figure 3a, revealing the main and side lobes. Generally, the narrower the main lobe and the shorter the side lobes the better the array and processing algorithm are at accurately identifying the correct $\vec{k}$ values for a given frequency.

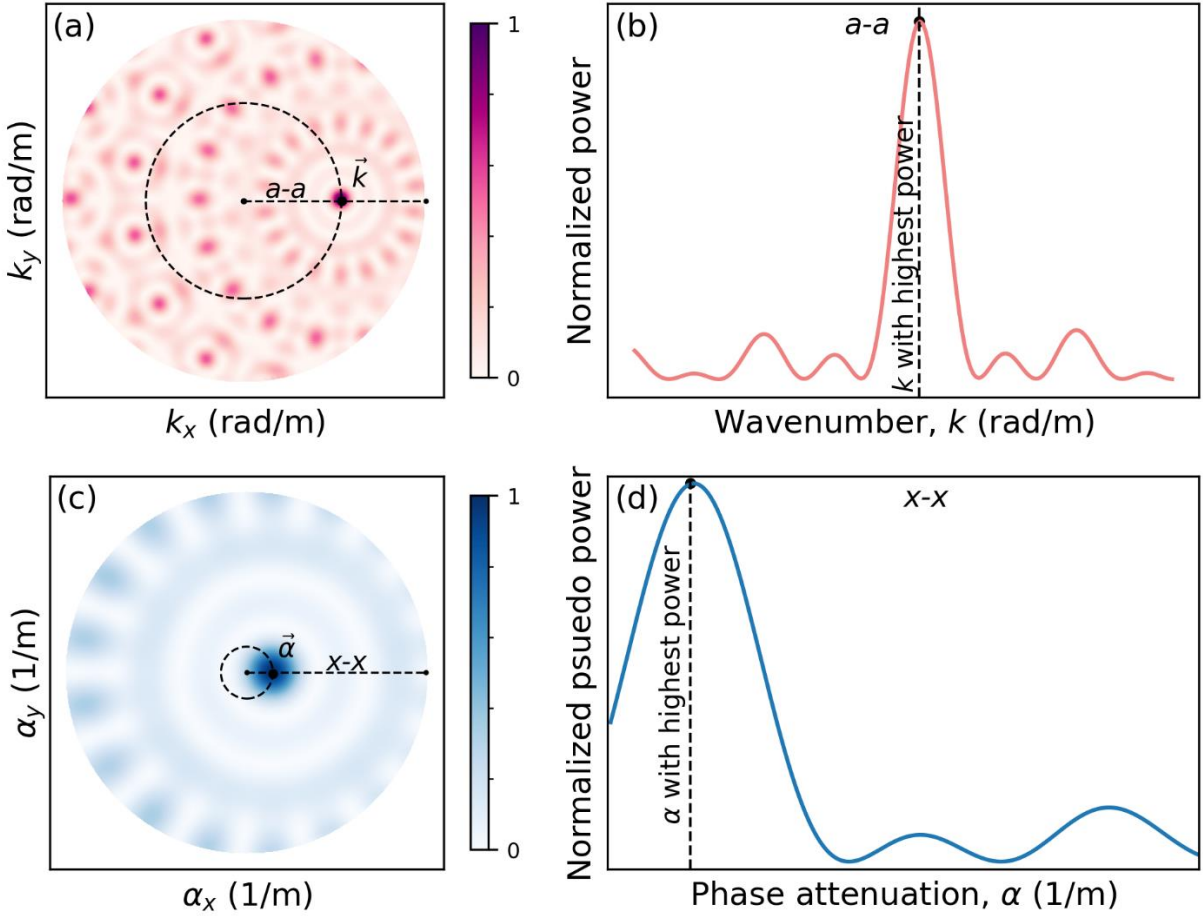

**Figure 3.** Schematic illustrating the FDBF and NFDBFa responses obtained from an ambient noise wavefield recorded by a ten-receiver circular MAM array for a single frequency and single time window. Panel (a) presents the $f$-$\vec{k}$ spectrum resulting from applying the FDBF method to the original wavefield, displaying the beamforming peak powers in $k_x$-$k_y$ space. Panel (b) shows the cross-section a-a from Figure 3a, revealing the main and side lobes. Panel (c) presents the $f$-$\vec{\alpha}$ spectrum resulting from applying the NFDBFa technique to the pseudo wavefield, presenting the beamforming peak powers in $\alpha_x$-$\alpha_y$ space. Panel (d) illustrates the cross-section x-x from Figure 3c, showing the main and side lobes along the direction of wave propagation.

Figures 3c and 3d display the $f$-$\vec{\alpha}$ spectrum obtained from applying the NFDBFa method to the converted noise wavefield for the same time window used to develop Figure 3a. In this case, instead of presenting the beamforming peak powers in the $k_x$-$k_y$ space, as seen in Figure 3a, they are now depicted in the $\alpha_x$-$\alpha_y$ space. This transition occurs because the phase in the pseudo wavefield is modulated by $\alpha$ (refer to Figure 2), rather than $k$. Figure 3c employs a different color scheme, where stronger powers are represented by darker blue colors. The $f$-$\vec{\alpha}$ spectrum shown in Figure 3c illustrates wave propagation for a single frequency along the x-axis with a phase attenuation represented by the vector $\vec{\alpha}$. Figure 3d illustrates the cross-section x-x from Figure 3c, revealing the main and side lobes along the positive x-axis (i.e., direction of wave propagation). Similar to estimating $\vec{k}$, the narrower the main lobe and the shorter the side lobes the better the array and processing algorithm are at accurately identifying the correct $\vec{\alpha}$ values for a given frequency. The ability of the NFDBFa approach to develop phase attenuation estimates from ambient-noise recorded using MAM arrays is investigated in the following section using synthetic data.

**NFDBFa evaluation with synthetic wavefields**

This section uses synthetic data to validate the effectiveness of the NFDBFa approach in estimating phase attenuation from ambient noise recorded using MAM arrays. Specifically, the approach is tested on two soil models: a half-space model and a single layer above a half-space model. All numerical simulations discussed in this section were executed using Salvus (Afanasiev et al., 2019), a comprehensive 2D and 3D full-waveform modeling software suite based on the spectral element method. The simulations were performed on the Texas Advanced Computing Center's (TACCs) high-performance cluster Lonestar6 using two compute nodes.

*Half-space model*

This subsection presents a simple wave propagation simulation consisting of a single surface source generating body and surface waves propagating through a half-space soil model. Despite the simplicity of the model, the outcomes obtained from this simulation offer key insights into the attenuation of a wavefield generated by a surface source and elucidate the capabilities of the NFDBFa approach. Figure 4 depicts a schematic plan view illustrating the source location and MAM array configurations employed in the half-space simulation. The wavefield was generated by a point source acting in the vertical direction at coordinates (0, 0, 0) in an x, y, z cartesian coordinate system. The source was a single Ricker wavelet with a center frequency of 5 Hz. This source function produces broadband energy over a frequency range of approximately 1 to 10 Hz. The wavefield emanating from the source was recorded using five circular MAM arrays, each comprising 10 sensors, with one sensor at the center and nine sensors evenly spaced around the perimeter. In this paper, the arrays are named using the convention 'C' followed by the diameter of the array, where 'C' denotes that the array is circular. Therefore, the first array, located two kilometers away from the source and with a diameter of one kilometer, is denoted as C1000 at 2 km. The remaining four arrays, concentrically-centered five kilometers from the source, have diameters of 60 m (C60), 300 m (C300), 1000 m (C1000 at 5 km), and 2000 m (C2000). It is noteworthy that, although currently only the vertical component of the displacement wavefield is utilized in NFDBFa, each sensor recorded both horizontal and vertical displacement components, and plans for utilizing all components from noise recordings are ongoing. Additionally, the NFDBFa processing operated independently of any knowledge about the source location, mirroring the conditions of an ambient noise MAM survey, ensuring an unbiased analysis.

The half-space constitutive soil parameters are presented in Figure 5a, where $V_p$, and $v$ are the compression wave velocity, and Poisson's ratio, respectively. Due to the large spatial extent of the model and the substantial computational expense associated with running a simulation over such a vast domain, a 2D simulation was conducted rather than a 3D simulation. In the 2D simulations, the sensor locations were projected onto a 2D plane, as illustrated in Figure 5a. This entailed setting the y-coordinate to zero for each surface sensor location shown in Figure 4, resulting in their positions being determined exclusively by their x-axis coordinates. For example, the sensor initially situated at coordinates (2321.4, 383, 0) in an x, y, z system (as depicted in Figure 4), transformed to (2321.4, 0) in the 2D x, z system presented in Figure 5. However, it is important to note that, during NFDBFa processing, the coordinates assigned to each sensor were derived from those shown in Figure 4; consequently, the aforementioned sensor retained coordinates of (2321.4, 383, 0) during processing. This approach not only substantially reduced the computational cost of the simulations, but also ensured that the arrays were measuring plane waves. The simulation required 4 hours and 20 minutes of computation utilizing 256 threads on the high-performance cluster Lonestar6.

Before describing the application of the NFDBFa method, some preliminary features of the amplitude decay versus distance are discussed, as they directly influence attenuation estimates. To better observe this decay pattern, the wavefield emanating from the source was recorded every ten meters along the surface. Those time histories were then filtered at discrete frequencies so the amplitude decay at each frequency could be observed. The decay of Fourier amplitudes with distance from the vertical Ricker wavelet source for frequencies 1, 2, 3, 4, and 5 Hz are shown in Figures 5b and 5c. In Figure 5b, the amplitudes for each frequency are normalized by their respective maximum values at the source and plotted on a log scale, while the distances are not normalized and plotted on a linear scale. In contrast, in Figure 5c, the distances from the source are normalized by the wavelength ($\lambda$) corresponding to each plane wave frequency and plotted on a linear scale. The figures depict a sharp amplitude decrease near the source due to nearfield effects. Following this, amplitude oscillations with diminishing power are superimposed over a linear decay trend. Note that a linear decay trend in log amplitude scale corresponds to an exponential decay in linear amplitude scale. These amplitude oscillations tend to flatten greatly after propagating approximately 10 $\lambda$ away from the source. It is noteworthy that these oscillations, although verified using other software packages, such as the ElastoDynamics Toolbox (EDT; Schevenels et al., 2009), challenge conventional intuition regarding wave attenuation in a half-space. Neither the geometric spreading of Rayleigh waves nor the attenuation due to material damping should exhibit such oscillations in a half-space, as detailed by Lai (1998). The oscillating amplitude decay pattern in a half-space model is a result of body wave amplitude decay oscillations, as shown by Tokimatsu (1995) and Holzlohner (1980). Hence, when estimating phase attenuation using ambient noise, it is essential for the MAM arrays to be at a sufficient distance (more than approximately 10 $\lambda$) away from any potential surface sources, such that wave amplitude oscillations do not contaminate the expected trend of amplitude decay with distance.

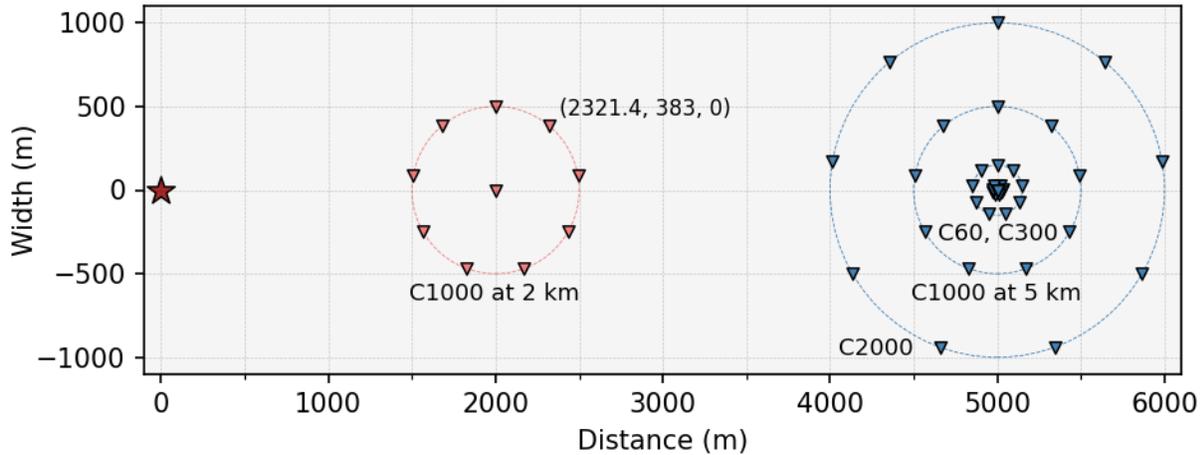

**Figure 4.** Plan view of the source (star symbol) and receiver (inverted triangle symbols) configurations used for synthetic wavefield simulations. The source was a single Ricker wavelet with a center frequency of 5 Hz. The wavefield was recorded using five MAM arrays. The first array (C1000 at 2 km) has a diameter of 1 km and is positioned 2 km from the source. The remaining four arrays are concentrically-centered 5 km away from the source and have diameters of 60 m (C60), 300 m (C300), 1 km (C1000 at 5 km), and 2 km (C2000), respectively.

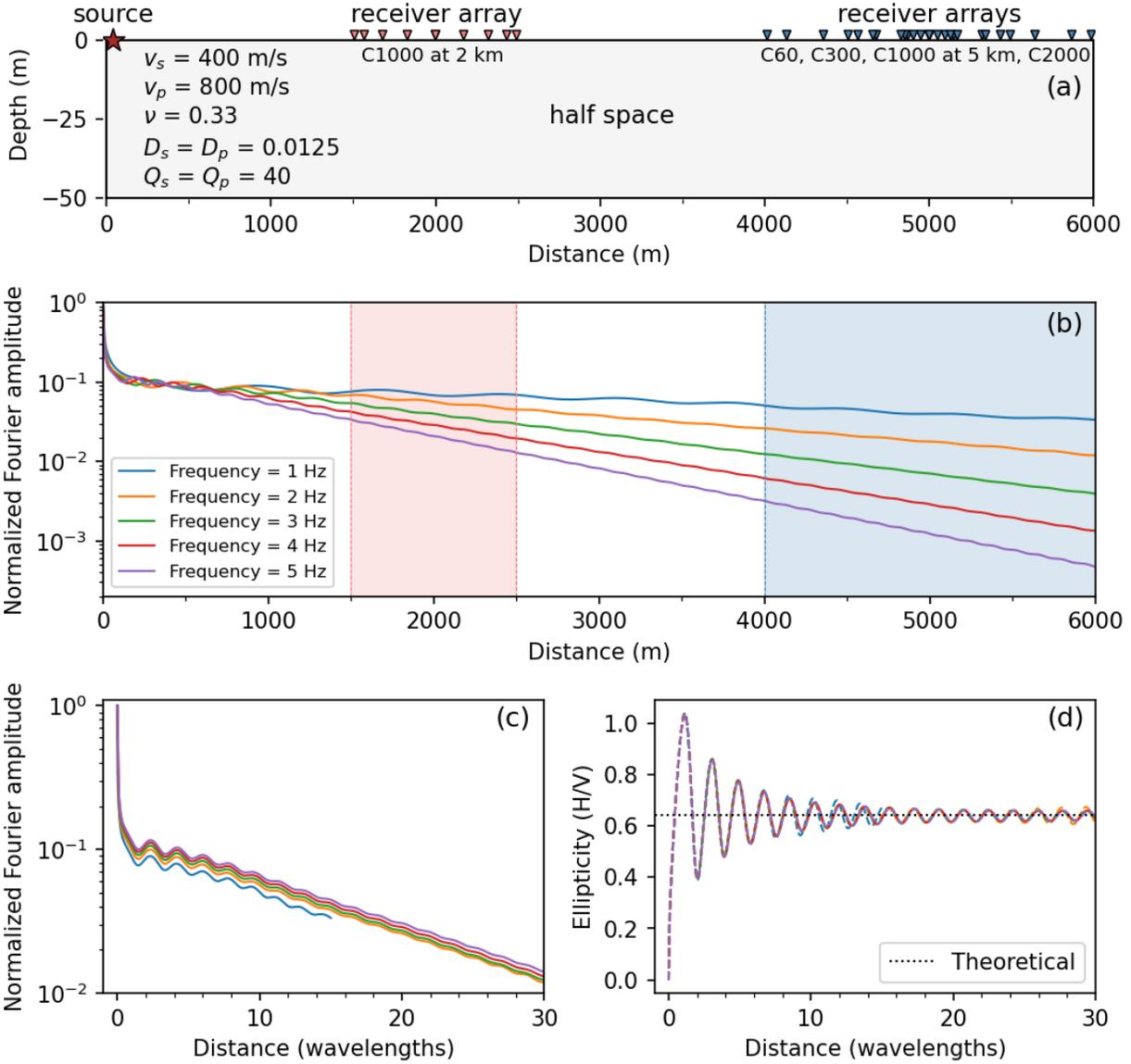

**Figure 5.** Half-space wavefield simulation: Panel (a) presents a cross-section view of the configuration of the source and receivers shown in Figure 4, along with the half-space soil properties. Panel (b) shows the decay of particle vertical displacement as a function of distance from the source for five distinct frequencies, each normalized by its maximum amplitude at the source. Panel (c) presents the particle displacement decay patterns from Panel b, with distance now normalized by the wavelength for each frequency. Panel (d) shows the particle ellipticities for each frequency, expressed as the horizontal particle displacement divided by the vertical particle displacement, with the dotted horizontal line indicating the theoretical ellipticity calculated based on the Poisson's ratio of the half-space soil model.

It is worth noting that in layered media, oscillating amplitude decay of Rayleigh waves due to geometric spreading has been reported and accounted for in attenuation studies, as observed in the work of Lai (1998). Thus, in layered media, wave amplitude oscillations can be more pronounced and may extend beyond 10 $\lambda$ from the surface source, as demonstrated by Tokimatsu (1995). This may be thought of as a type of near-field effect specific to attenuation studies, wherein the wavefield amplitude decay patterns are significantly more complicated at distances less than approximately 10 $\lambda$ from source. This is distinct from, and more severe than, the typical range of near-field effects for phase velocity estimations, which generally deteriorate between 0.5 $\lambda$ and 2 $\lambda$ from the source, depending on the subsurface velocity structure (Tokimatsu 1995; Rix et al., 2001).

To further demonstrate the more severe near-field effects associated with amplitude decay, Figure 5d presents the simulated wavefield ellipticity, expressed through the horizontal-to-vertical (H/V) ratio of particle displacement, measured with distance in wavelengths for the same frequencies outlined in Figure 5b. The ellipticity also displays oscillations that decrease and stabilize at normalized distances greater than about 10 $\lambda$ from the source. This observation underscores that the near-field amplitude decay oscillations stem from body waves, as Rayleigh wave ellipticity in a half-space is determined solely by Poisson's ratio (Tokimatsu 1995) and should not oscillate. In Figure 5d, we observe that the calculated ellipticities oscillate around the theoretical value anticipated for Rayleigh wave ellipticity in a half-space with Poisson's ratio equal to 0.33, depicted by the dotted horizontal line in Figure 5d.

The synthetic time histories recorded by the C1000 at 2 km and the C1000 at 5 km MAM arrays (refer to Figures 4 and 5) were processed using the FDBF and NFDBFa methods to estimate phase velocity and attenuation, respectively, as illustrated in Figure 6. Figure 6 aims to highlight the impact of wave amplitude decay patterns on the attenuation estimates. In terms of abilities to resolve phase velocity, both the C1000 arrays seem to perform approximately the same, whether 2 km away from the source (Figure 6a) or 5 km away from the source (Figure 6b). However, upon inspecting Figures 6c and 6d, it becomes evident that the array located 5 km from the source (i.e., Figure 6d) provides more reliable attenuation estimates at lower frequencies compared to the array closer to the source. This observation can be explained by referring to Figure 5b, where the amplitude decay patterns measured by the array positioned 2 km from the source are shaded in pink. It is apparent that in close proximity to the source, the low-frequency waves have not traveled a sufficient number of wavelengths, resulting in amplitude decay that does not conform to pure exponentials (i.e., linear decay in log scale). However, by the time these waves reach the array positioned 5 km from the source (blue shading in Figure 5b), the oscillations in amplitude decay have diminished significantly, approaching a pure exponential decay. Therefore, it is noteworthy that in an ambient-noise survey, even though the source location is unknown, if the noise source is close to the array in terms of wavelengths traveled by the desired frequency, it may lead to unreliable and scattered attenuation results. Nonetheless, Figures 6c and 6d clearly demonstrate the reliability of the new NFDBFa approach in retrieving phase attenuation estimates over a broad range of frequencies.

Finally, the performance of the NFDBFa in the presence of incoherent noise is investigated. For this purpose, Figure 7 illustrates the influence of incoherent noise and array size on phase attenuation estimates using the same half-space simulation results. The analysis focuses on the four arrays of different sizes concentrically-centered 5 km from the source (refer to Figures 4 and 5a). Incoherent noise was introduced to the signal, with a target signal-to-noise ratio (SNR) at 20 dB, which resulted in the frequency-dependent amplitude decay patterns depicted in Figure 7a (compared to Figure 5b). Figures 7b, 7c, 7d, and 7e display the attenuation estimates obtained using the C60, C300, C1000, and C2000 MAM arrays, respectively. It becomes evident that larger arrays yield more accurate attenuation estimates in the presence of incoherent noise. Figure 7a elucidates the rationale behind this enhanced performance for larger arrays across all frequencies. The C2000 MAM array samples a significantly larger area, enabling it to discern the exponential amplitude decay even in the presence of noise. The C60 MAM array samples a significantly smaller area, and thus is considerably more sensitive to amplitude fluctuations caused by incoherent noise, resulting in the significant scatter observed in the attenuation estimates shown in Figure 7b.

Figure 8 further illustrates the impact of array size on resolving attenuation coefficients by showcasing the $f$-$\vec{\alpha}$ spectra for a frequency of 3 Hz that were calculated from the wavefield recorded by the four concentrically-centered arrays located 5 km from the source. Notably, the mainlobe (dark blue shaded area) is considerably narrower for larger arrays, resulting in more reliable estimates of phase attenuation. Two key points warrant attention here. First, the feasibility of employing larger arrays might be restricted due to limitations in access at a given site, or to help maintain approximately a one-dimensional (1D) subsurface condition beneath the array, which is an implicit assumption in the analysis technique (i.e., no lateral spatial variability). Meeting this assumption becomes more challenging as the array size expands. Second, it is essential to highlight that the method used to determine the optimal MAM array size for attenuation estimates differs from the one employed in obtaining phase velocity estimates. In dispersion estimation, smaller MAM arrays are more effective at capturing high frequency phase velocities, whereas larger arrays are better suited for resolving lower frequency phase velocities (Foti et al., 2018; Vantassel and Cox, 2022). However, according to the results depicted in Figure 7, the larger arrays demonstrated superior ability in resolving phase attenuation across the entire considered frequency range compared to the smaller arrays.

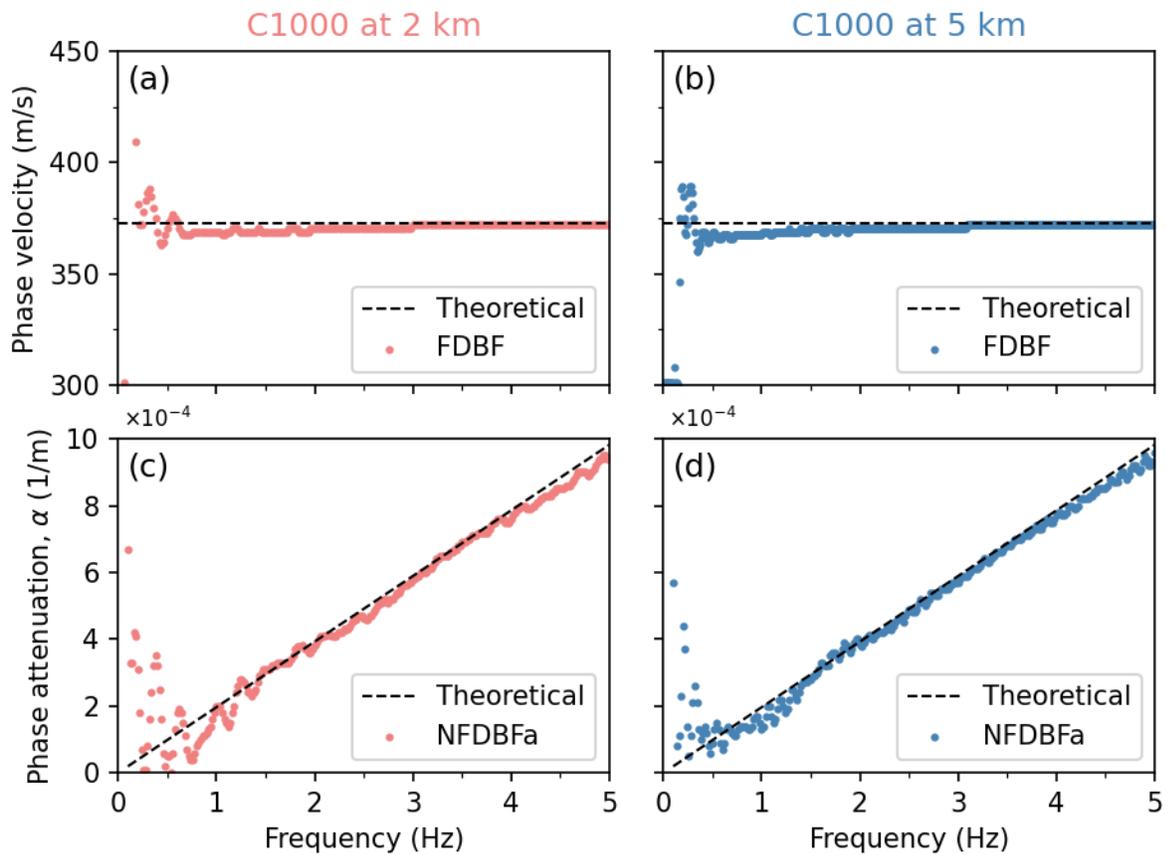

**Figure 6.** Half-space wavefield simulation: phase velocity (top) and phase attenuation (bottom) dispersion data estimated with FDBF and NFDBFa, respectively, from 1 km arrays positioned at two distinct distances from the ambient noise source: (left) at two kilometers (C1000 at 2 km), and (right) at five kilometers (C1000 at 5 km).

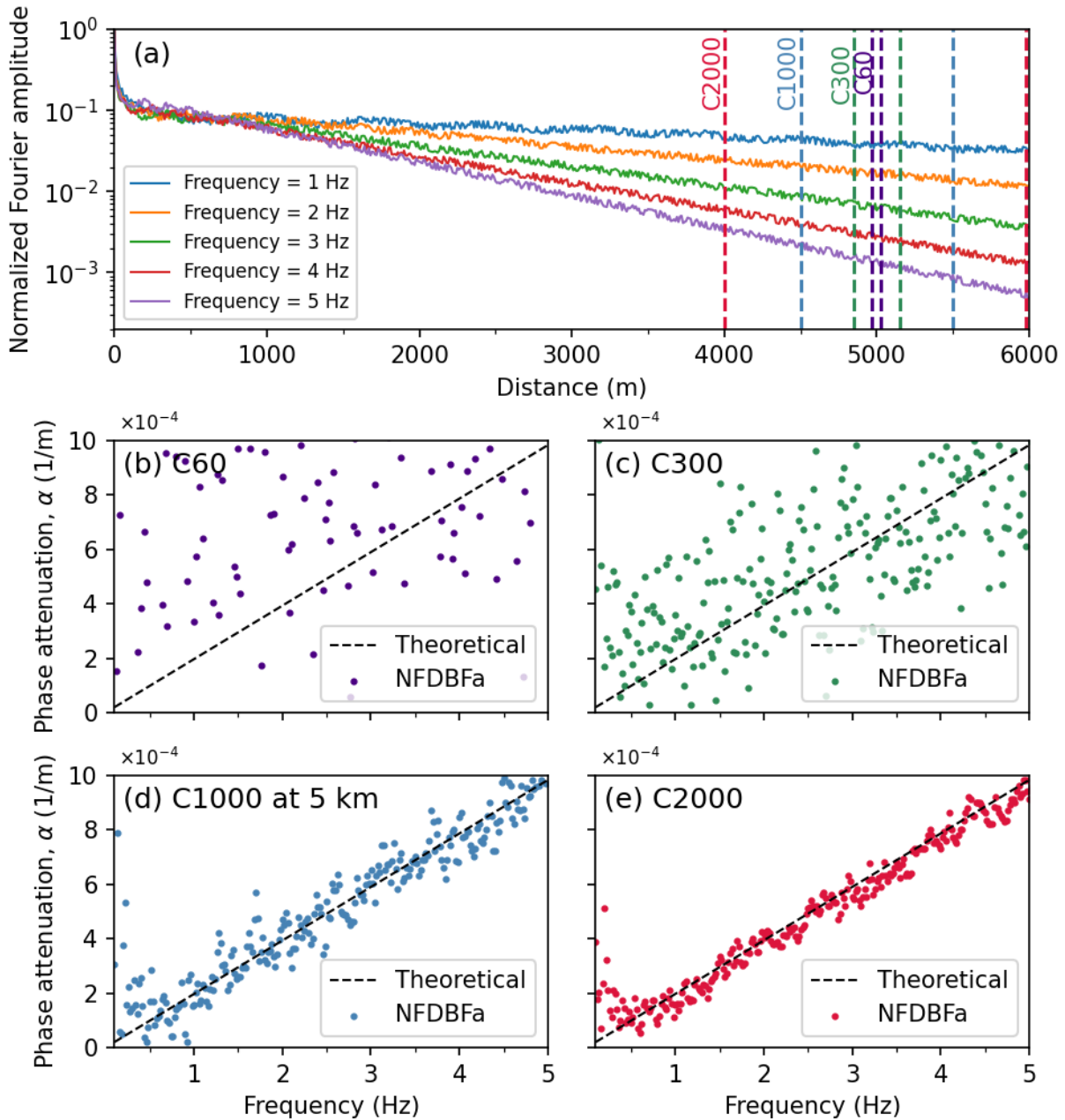

**Figure 7.** Half-space wavefield simulation with noise: Panel (a) shows the amplitude decay of the same five frequencies depicted in Figure 5 but now with added incoherent noise to the signal, setting the signal-to-noise ratio (SNR) at 20 dB. Panels (b) to (e) present the predicted phase attenuation data from the NFDBFa analysis for four arrays concentrically-centered at five kilometers from the source, with diameters of 60 m (C60), 300 m (C300), 1 km (C1000 at 5 km), and 2 km (C2000), respectively.

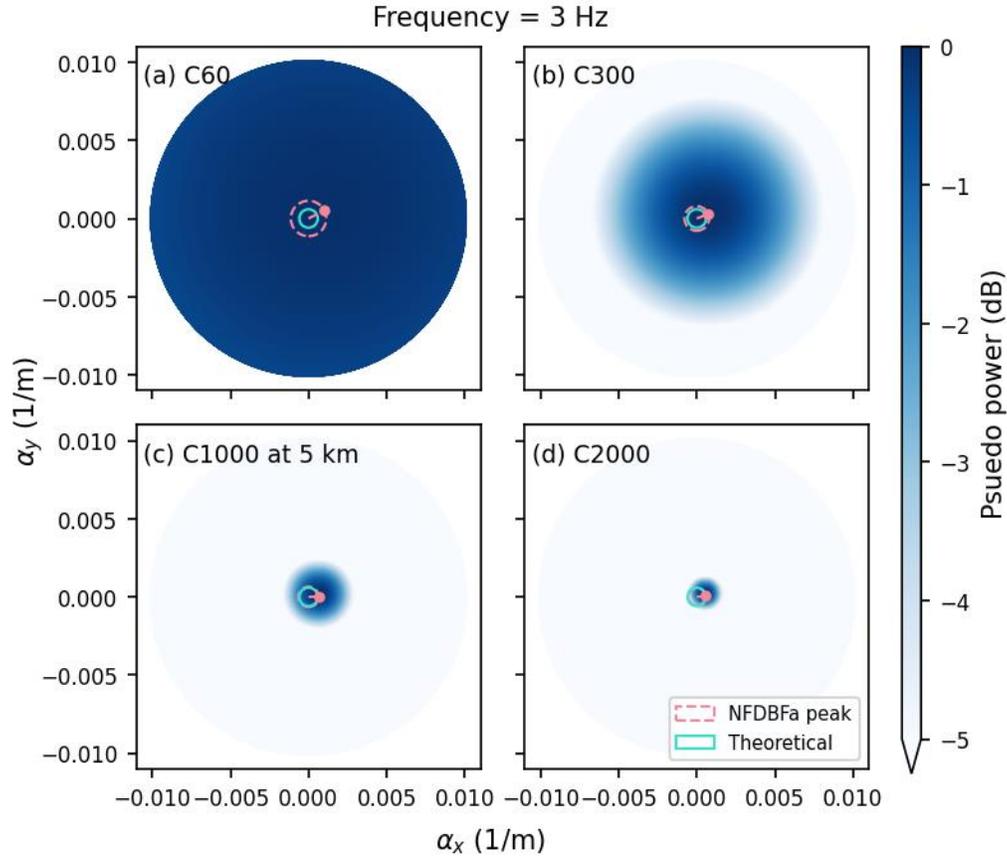

**Figure 8.** Half-space wavefield simulation with noise: Panels (a) through (d) present the *f*-$\vec{\alpha}$ spectra obtained through NFDBFa analysis for a frequency of 3 Hz. The spectra are derived from the wavefield recorded by the four arrays concentrically-centered five kilometers from the source with diameters of 60 m (C60), 300 m (C300), 1 km (C1000), and 2 km (C2000), respectively, as depicted in Figure 7.

### *Layer above a half-space model.*

The performance of the NFDBFa approach on a synthetic model consisting of a single layer above a half-space is illustrated in this subsection. The model's constitutive small-strain parameters and the source and receiver configurations are provided in Figure 9a. For this synthetic study, 150 vertical point sources with varying forcing functions and trigger times were activated. The sources were triggered five kilometers away from the center of a one-kilometer diameter circular array consisting of 10 sensors: one in the center and nine equally distributed around its perimeter (just like the C1000 at 5 km MAM array depicted in Figure 4). The waveforms recorded by the array are depicted in Figure 9b. These waveforms were subsequently processed using FDBF and NFDBFa to derive the Rayleigh wave phase velocity dispersion data shown in Figure 9c and the phase-attenuation data shown in Figure 9d, respectively. The theoretical Rayleigh-wave phase velocity dispersion and attenuation curves for the model are also presented in Figures 9c and 9d, respectively. In these figures, the fundamental theoretical mode is denoted as Mode 1, while the 1st-higher mode is denoted as Mode 2.

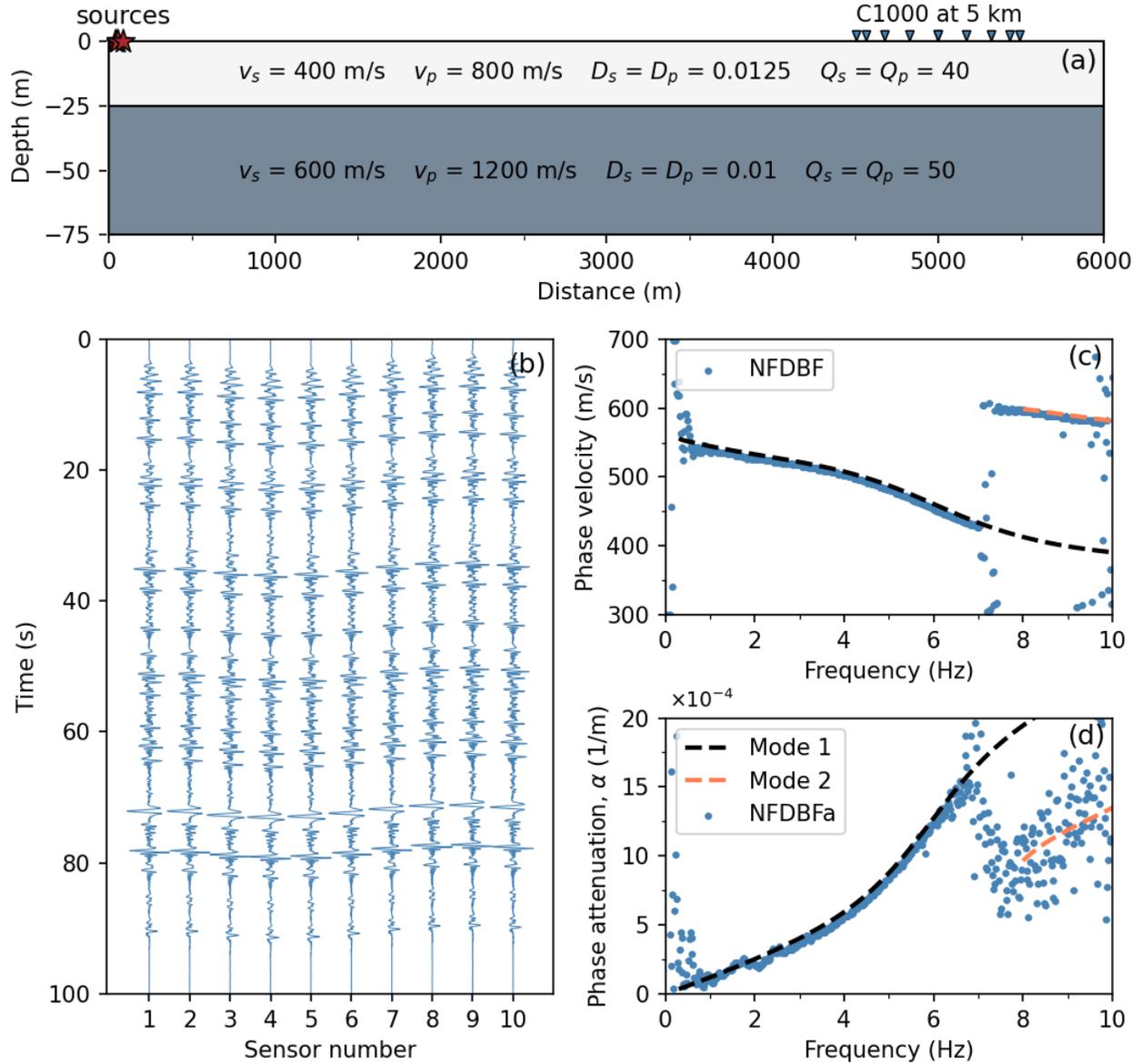

**Figure 9.** Layered model simulation: Panel (a) presents the soil properties utilized in the simulation for the soil layer and the half-space, along with the surface sources and 1-km receiver array located 5 km from the source (C1000 at 5 km). Panel (b) displays the waveforms collected from the C1000 array. In Panel (c), the good agreement between the theoretical Rayleigh-wave phase velocity curves (Mode 1 and Mode 2) and the experimental phase velocity data obtained through the FDBF approach on the original wavefield is demonstrated. Finally, Panel (d) showcases the good agreement between the theoretical phase attenuation curves and the experimental phase attenuation data extracted from the converted wavefield using the proposed NFDBFa approach are depicted.

The FDBF method is able to extract experimental phase velocity dispersion data from the synthetic wavefield that well-matches the theoretical dispersion curves and captures the transition from Mode 1 to Mode 2 at approximately 7 Hz. A strong agreement is also observed between the theoretical attenuation curves and the experimental attenuation data extracted from the synthetic wavefield using the NFDBFa method, particularly for Mode 1. Interestingly, the attenuation data shifts to Mode 2 at the same frequency where the phase velocity dispersion data transitions to Mode 2. A similar observation about possible links between the frequencies where phase velocity and attenuation mode transitions occur was also reported by Aimar et al., (2024a) using MASW data. While further studies are needed to validate the observations that phase velocity and phase attenuation data tend to jump modes at identical frequencies, this is a potentially important point, as patterns in attenuation modes are much more complex than phase velocity modes.

The effectiveness of the proposed NFDFBa approach has been successfully demonstrated through the analyses conducted on synthetic datasets, as discussed above. Now, we shift our focus to applying this approach to real field data, offering a thorough demonstration of its effectiveness in a practical, real-world situation.

**Field application and validation**

A surface wave field-testing campaign was conducted at the Drainage Farm Site in Logan, Utah, USA (refer to Figure 10), a property owned by Utah State University (USU). Structural geology indicates that Southern Cache Valley, encompassing the Drainage Farm Site and located in the northeastern part of the Basin and Range province, is a graben bounded by high-angle normal faults (Williams, 1962). The site is underlain by Paleozoic rocks, which are overlain by Tertiary formations such as the Wasatch and Salt Lake formations, composed of conglomerate, siltstone, and tuffaceous sandstone. In certain areas of Cache Valley, these formations reach thicknesses of up to 2,440 m (Evans et al., 1996). The near-surface geology of the Drainage Farm Site is characterized by sediments from ancient Lake Bonneville, which receded to form the Provo shoreline. These sediments include alluvial, lacustrine, and deltaic deposits (Williams, 1962; Evans et al., 1996). Well logs presented by Williams (1962) reveal alternating layers of silt and clay, sand, and gravel above the Salt Lake formation. Moreover, limited deep well logs from the vicinity of the Drainage Farm Site indicate that rock can be encountered at depths ranging from 176 m to more than 350 m (Perez, 1969).

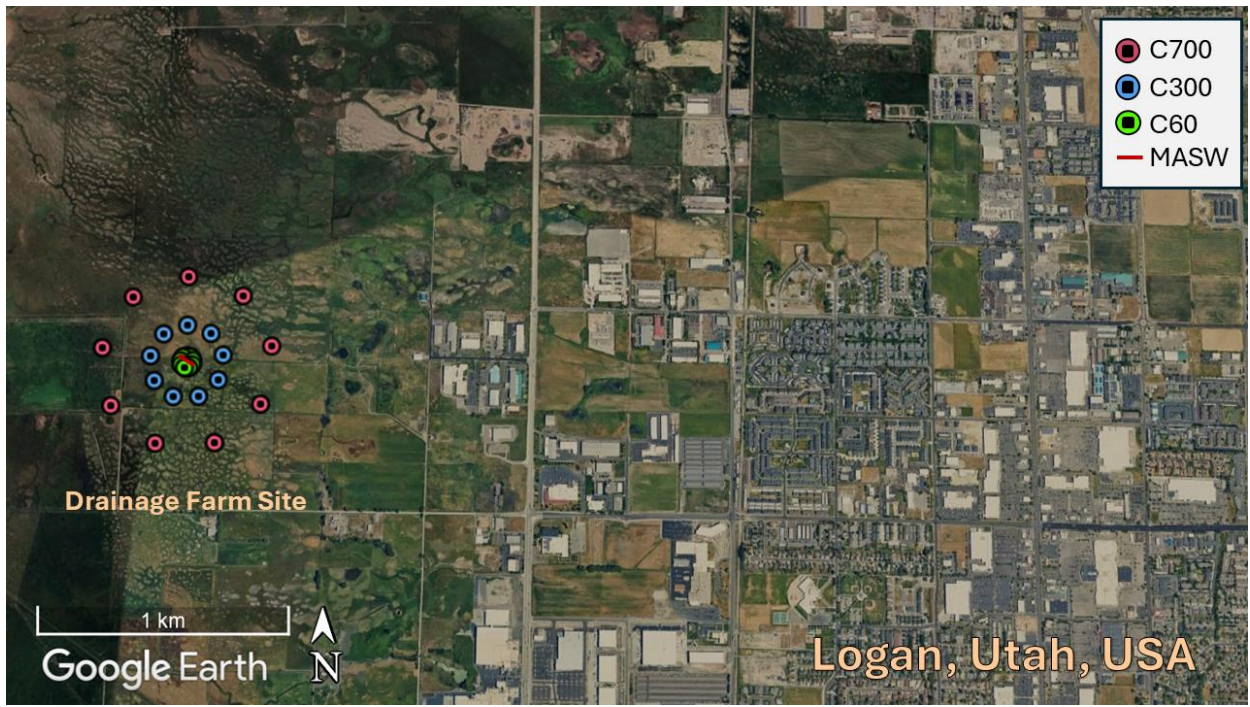

**Figure 10.** Plan view of the MASW and MAM arrays employed for testing at the Drainage Farm Site in Logan, Utah, USA. The concentric MAM arrays featured diameters of 60 m (C60), 300 m (C300), and 700 m (C700), while the MASW array comprised 24, 4.5-Hz vertical geophones, spanning 46 m.

The goal of the testing was to collect a high-quality surface wave dataset that could be used for attenuation studies to validate the proposed NFDBFa technique. The field testing involved both active-source MASW testing and ambient noise MAM testing. The sensor array configurations utilized for MASW and MAM at the Drainage Farm Site are illustrated in Figure 10. MASW testing was performed using 24, 4.5-Hz vertical geophones placed with a spacing of two meters between successive geophones, resulting in an array length of 46 m. Wavefields with strong Rayleigh wave content were actively generated by striking vertically on a strike-plate with a sledgehammer. The sledgehammer was used at eight distinct "shot" locations that were offset by 5, 10, 15, and 20 m relative to the first/last geophone off each end of the array. Five distinct sledgehammer blows were recorded at each location for subsequent stacking to increase the signal-to-noise ratio (Foti et al. 2018). MAM testing utilized three concentric circular arrays that were aligned with the middle of the MASW array, as depicted in Figure 10. The three arrays were 700-m, 300-m, and 60-m in diameter, and will be referred to as C700, C300, and C60, respectively. Each array consisted of nine evenly distributed three-component broadband seismometers (Nanometrics Inc. Trillium Compact 120s seismometers) along its circumference to capture ambient vibrations. The three arrays did not record data simultaneously; instead, the nine sensors were used to collect noise data for each of the MAM arrays one array at a time. First, the sensors recorded seismic noise for 13 hours and 30 minutes for the C700 array. Subsequently, the sensors were relocated to their designated locations for the C60 and C300 arrays, recording ambient noise for an hour and a half and three hours, respectively.

For Rayleigh-wave phase velocity dispersion analysis, MASW data were analyzed using the FDBF method with cylindrical-wave steering (Zywicki and Rix, 2005), as coded in the open-source surface wave processing package swprocess (Vantassel, 2021). This processing was coupled with the multiple source-offset technique for identifying near-field contamination and quantifying dispersion uncertainty (Cox and Wood, 2011; Vantassel and Cox, 2022). As a result, eight phase velocity estimates were obtained for each frequency, corresponding to one phase velocity estimate from each of the eight shot locations. MASW Rayleigh wave dispersion data influenced by near-field effects or significant offline noise were trimmed before calculating phase velocity dispersion statistics.

The three-component beamforming approach (Wathelet et al., 2018) coded in the open-source software package Geopsy (Wathelet et al., 2020) was used to generate Rayleigh-wave phase velocity dispersion data for each of the MAM arrays. The recorded time for each array was discretized into blocks, with each block further divided into at least 30-time windows. The window lengths were selected to contain at least 30 cycles (periods) at the lowest processing frequency that could be extracted from each MAM array (Vantassel and Cox, 2022). For each MAM array, eight phase velocity estimates were extracted at each analyzed frequency using the three-component beamforming (Wathelet et al., 2018) approach to ensure consistency with the eight phase velocity estimates obtained from the MASW processing. Spurious dispersion data stemming from high-amplitude noise in the near-field (e.g., traffic noise close to the sensors) and incoherent noise were manually eliminated before calculating dispersion statistics. Ambient noise phase velocity dispersion data from all MAM arrays were combined with the active phase velocity dispersion data obtained from MASW processing, as shown in Figure 11a. The combined data, used to compute mean and ± one standard deviation dispersion estimates (Vantassel and Cox, 2022), are displayed in Figure 11b relative to the individual MASW and MAM dispersion data points for the Drainage Farm Site.

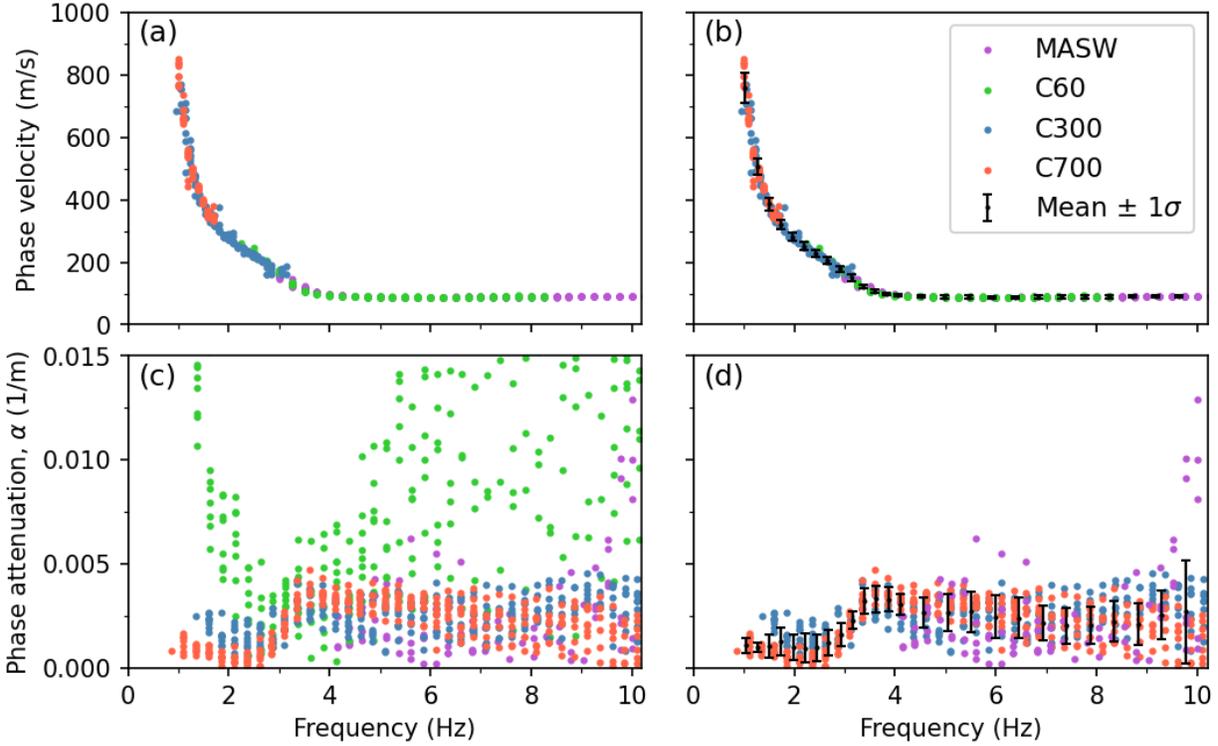

**Figure 11.** Experimental phase velocity and attenuation data extracted from MASW and MAM testing at the Drainage Farm Site in Logan, UT, USA. Panel (a) displays the experimental phase velocity dispersion data of Rayleigh waves processed from an MASW array and three circular MAM arrays, with diameters of 60 m, 300 m, and 700 m. Panel (b) showcases the mean and ± one standard deviation of the experimental Rayleigh wave phase velocity dispersion data derived from the combined MASW and MAM datasets. Panel (c) displays the experimental phase attenuation data from MASW and three circular MAM arrays. Panel (d) illustrates the mean ± one standard deviation of the experimental phase attenuation data calculated from the combined MASW, C300, and C700 MAM arrays.

The cylindrical FDBFa (CFDBFa) approach, as proposed by Aimar et al. (2024a), was employed to derive attenuation estimates from the MASW data. Mirroring the MASW phase velocity dispersion analysis, the multiple source-offset technique was utilized for quantifying attenuation uncertainty. Thus, eight attenuation estimates were extracted from the MASW data at each analyzed frequency using CFDBFa. For the MAM attenuation estimates, the new NFDBFa approach introduced in this study was employed. The recorded time for each array was discretized into eight blocks, with each block further divided into 30 windows. Consequently, the window length employed for each MAM array can be determined by dividing the total recording time of the array by the product of 8 blocks and 30 windows (i.e., 240). Similar to MAM phase velocity dispersion analysis, the window lengths were selected to contain at least 30 periods at the lowest processing frequency that could be extracted from each MAM array (Vantassel and Cox, 2022). Averaging the estimates from all windows within each block yielded a single data point per block, thus providing eight unique attenuation estimates per frequency. This processing approach ensured that an equal number of attenuation data points were obtained at each frequency for all of the MASW and MAM arrays. The combined ambient-noise attenuation data from all MAM arrays and the active attenuation data from the MASW array are plotted together in Figure 11c. A good agreement is observed between the attenuation estimates derived from the MASW array and those obtained from the C300 and C700 arrays for frequencies ranging from 4 to 10 Hz. The MASW testing did not generate coherent attenuation data at frequencies less than 4 Hz, due to the limitations of the active sledgehammer source. However, the MAM testing was able to extract coherent attenuation data at frequencies below 1 Hz. The agreement observed between the active-source and ambient noise attenuation estimates serves as compelling evidence for the efficacy of the proposed NFDBFa approach. However, it is notable that there is significant scatter in the attenuation estimates obtained using the C60 array. This variability is likely attributed to the challenges previously discussed in regards to using smaller MAM arrays for attenuation studies, as the phase velocity data extracted from the C60 array was very good (refer to Figure 11a). Hence, the attenuation data from the C60 array was removed prior to calculating attenuation statistics. The combined attenuation estimates from the MASW, C300, and C700 arrays, and the mean and ± one standard deviation attenuation estimates obtained from those three arrays, are depicted in Figure 11d. While a noticeable agreement exists among the three arrays, there is significantly greater scatter in the attenuation estimates (Figure 11d) compared to the phase velocity dispersion estimates (Figure 11c). This observation aligns with the findings reported by Aimar (2022). The application of the new NFDBFa approach in this field test showcases its effectiveness in estimating attenuation coefficients from ambient noise wavefield data.

Finally, the statistical experimental Rayleigh-wave phase velocity and attenuation parameters derived from both the MASW and MAM testing (refer to Figures 11b and 11d) were used to invert for $V_s$ and $D_s$ profiles at the Drainage Farm Site. This was achieved through the Monte Carlo-based joint inversion of phase velocity and phase attenuation data developed by Aimar et al. (2024b). Although the effectiveness of the joint inversion procedure has been proven for active surface wave data (Lai and Rix, 1998; Aimar et al., 2024b), its application to combined dispersion data from MASW and MAM testing, covering a broad frequency range, is novel. This is because past studies on inverting MAM-based attenuation data to retrieve damping properties at large depths typically adopted an uncoupled inversion approach, based on a separate inversion of Rayleigh-wave phase velocity and $\alpha$ (e.g., Prieto et al., 2009; Parolai, 2014).

The inversions performed herein involved 50,000 five-layer trial Earth models with progressively increasing thicknesses, covering a comprehensive range of layer thicknesses, $V_s$, and $D_s$ values. The layering was informed by a preliminary inversion study based solely on phase velocity dispersion data, which is omitted here for simplicity. Realistic values were fixed for the Poisson's ratio and mass densities. A constant $D_p/D_s$ ratio of 1.4 was used during the inversion, similar to the approach taken by Bergamo et al. (2023). Forward dispersion and attenuation modeling were conducted using the Computer Programs for Seismology software (Herrmann, 2013). The fit to the experimental data was quantitatively assessed using a normalized root mean square (RMS) error that accounts for estimation uncertainty, similar to the metric proposed by Wathelet et al. (2004).

The ten best inversion results are shown in Figure 12. The theoretical phase velocity and attenuation curves are shown relative to the experimental data in Figures 12a and 12b, respectively. The $V_s$ and $D_s$ profiles are shown in Figures 12c and 12d, respectively, down to a depth of 400 m, which is approximately 1/2 of the maximum resolved phase velocity wavelength. The $V_s$ profiles in Figure 12c collectively feature a shallow layer about 25 m thick with velocities ranging from approximately 90 to 185 m/s, including a low-velocity zone, which is consistent with known near-surface layering. Below this, there is generally a thicker layer extending down to approximately 180 m with velocities varying around 500 m/s. At depths of 150-200 m, a stiff layer with velocities around 1500 m/s is commonly identified across the profiles. These depths, while variable, are consistent with the location of Salt Lake Formation rock surface, as discussed above. The $D_s$ profiles in Figure 12d indicate that damping in the top 25 m is less than approximately 1%. Below this depth, there is a noticeable variability in the estimated $D_s$ values between the ten best profiles. Nonetheless, $D_s$ can be observed to increase to approximately 2% to 4% in the deeper soil deposits, which consist of alternating clay, sand, and gravel layers. At the top of the Salt Lake Formation rock surface, $D_s$ collectively decreases again to less than 2% for all of the ten best profiles. The large variability in $D_s$ can be attributed to the complex geology of the site, the significant standard deviation in the experimental attenuation data, and the moderately low sensitivity of theoretical attenuation curves to $D_s$ at greater depths (e.g., Badsar et al., 2012; Aimar et al., 2024b).

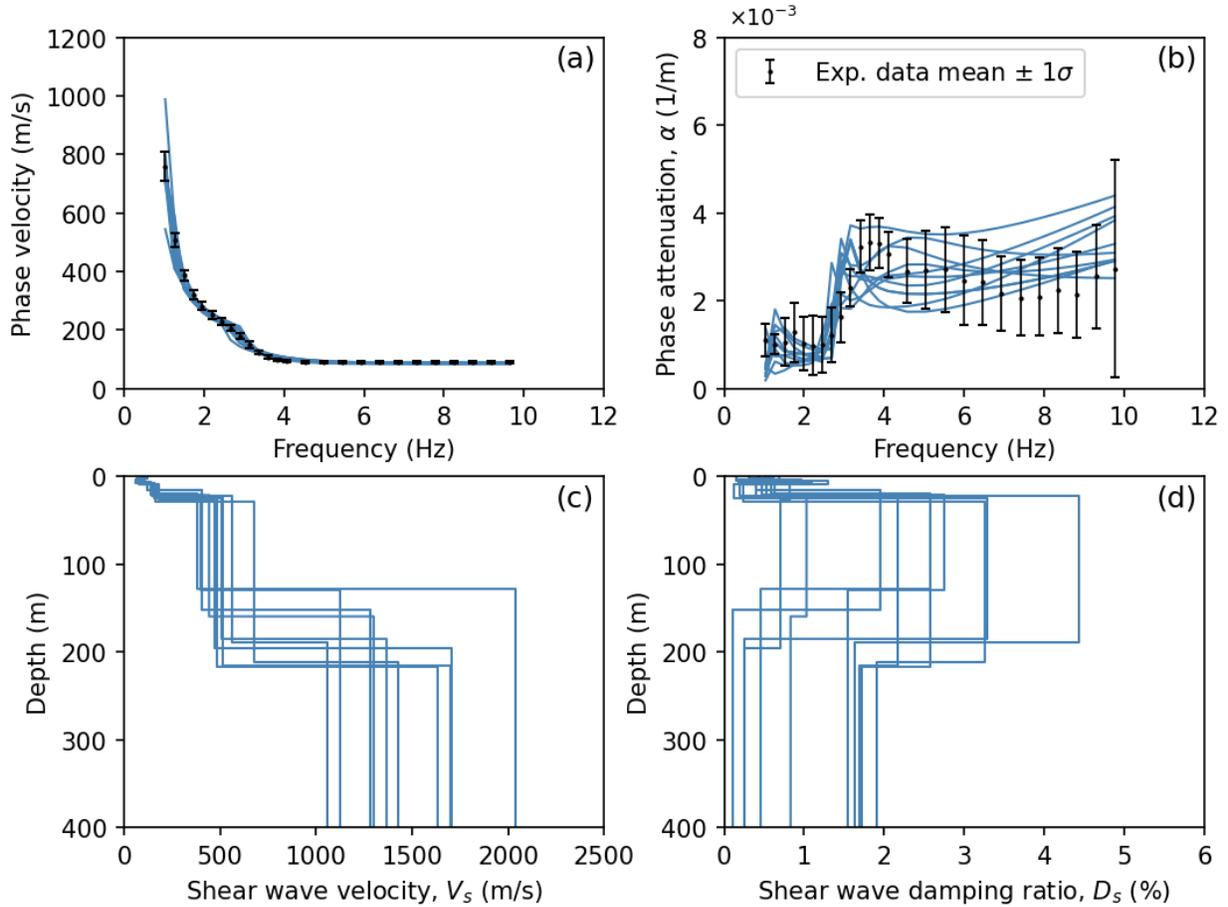

**Figure 12.** Inversion results for the experimental Rayleigh-wave phase velocity and attenuation data collected at the Drainage Farm Site in Logan, UT, USA. The figure highlights the ten best-fitting models, with Panels (a) and (b) comparing the theoretical curves for phase velocity and attenuation, respectively, against the experimental data represented by mean values with ± one standard deviation error bars. Panels (c) and (d) display the $V_s$ and $D_s$ profiles, respectively, for the ten best theoretical models.

Despite these limitations, the joint inversion procedure successfully provided in-situ estimates of $D_s$ at depths not reached by conventional site characterization techniques. This confirms the advantages of combining MASW and MAM data for the combined estimation of stiffness and dissipation parameters of soil deposits.

**Conclusions**

A new methodology for estimating frequency-dependent attenuation coefficients through the analysis of ambient noise wavefield data recorded by 2D arrays of surface seismic sensors has been presented. The approach relies on the application of an attenuation-specific wavefield conversion and frequency-domain beamforming (FDBF). It has been termed the noise FDBF attenuation (NFDBFa) method. Importantly, using an FDBF approach, as opposed to a noise cross-correlation approach, enables the direction of ambient noise propagation to be determined for each noise window and frequency, and does not require an equipartitioned ambient noise wavefield. Furthermore, using an FDBF approach enables the phase velocity and attenuation data generated from active-source testing like MASW to be combined with phase velocity and attenuation data generated from ambient noise testing like MAM in order to span a broader frequency range. This enables the joint inversion of phase velocity and attenuation to be performed as a means to extract shear wave velocity and small-strain damping ratio profiles to significantly greater depths than previously possible using only active-source data.

Numerical simulations were conducted to deepen our understanding of the proposed NFDBFa method. These simulations aimed to evaluate how the proximity of the MAM array to the noise source, the presence of incoherent noise, and the size of the array affect the estimates of phase attenuation. The results demonstrated that near-field effects are more pronounced and extend over greater distances for phase attenuation estimates in comparison to those considered for phase velocity estimation. Furthermore, it was discovered that larger array sizes consistently provided more accurate phase attenuation estimates across all considered frequencies, contrary to the conventional MAM design criteria used for phase velocity dispersion estimation, where larger arrays are typically preferred for resolving lower frequencies while smaller arrays excel at resolving higher frequencies. This distinction emphasizes the need for unique design criteria when planning a MAM array for attenuation estimation.

The proposed NFDBFa approach underwent validation through numerical wave propagation simulations, comparing predicted frequency-dependent phase attenuation values against theoretical phase attenuation curves for two synthetic models. Furthermore, validation of the developed technique was reinforced using MASW and MAM field data collected at the Drainage Farm Site in Logan, Utah, USA. The phase velocity and attenuation data extracted from the MASW and MAM recordings agreed well over a common bandwidth, while the ambient noise MAM data allowed the phase velocity and attenuation estimates to be extracted at significantly lower frequencies. The joint inversion of the experimental Rayleigh-wave phase velocity and phase attenuation data obtained from both MASW and MAM testing facilitated the estimation of shear wave velocity and small-strain damping ratio profiles to significant depths (400 m) at the Drainage Farm Site.

As noted herein and in other studies like Aimar et al. (2024a), attenuation data are significantly more variable and more complex to understand (e.g., modal curves that repeatedly cross one another) than phase velocity data. As such, there is a need for future studies to better understand attenuation data and how to invert them to retrieve reliable in-situ profiles of the small-strain damping ratio. Future efforts should involve additional numerical and experimental testing of diverse subsurface conditions, coupled with comparisons to damping estimates obtained from invasive tests. With the validity of this approach demonstrated on the vertical component, future research will also explore the utilization of the three components of the noise wavefield to enhance attenuation estimates beyond the current method's capabilities.

**Declaration of Competing Interest**

The authors declare that they have no known competing financial interests or personal relationships that could have appeared to influence the work reported in this paper.

**Acknowledgements**

The numerical simulations were run on the Texas Advanced Computing Center's (TACC's) cluster Lonestar6, with an allocation provided by DesignSafe-CI (Rathje et al., 2017). This work was supported by the U.S. National Science Foundation (NSF) Grant Number CMMI-2120155. However, any opinions, findings, conclusions, or recommendations expressed in this article are those of the authors and do not necessarily reflect the views of the NSF.

**Research Data and Code Availability**

The field test data used to validate the NFDBFa approach presented in this paper are available in the dataset by Abbas et al. (2024).

# References

Abbas A, Cox BR, Dawadi N, Jackson N, and Cannon K (2024) Geotechnical site characterization at the Drainage Farm Site. DesignSafe-CI. https://doi.org/10.17603/ds2-sx2h-8s20 v1

Aimar M, Foti S and Cox BR (2024a accepted) Novel Techniques for In-situ Estimation of Shear-wave Velocity and Damping Ratio through MASW testing Part I: A Beamforming Procedure for Extracting Rayleigh-wave Phase Velocity and Phase Attenuation. (accepted to Geophysical Journal International).

Aimar M, Foti S and Cox BR (2024b accepted) Novel Techniques for In-situ Estimation of Shear-wave Velocity and Damping Ratio through MASW testing Part II: A Monte Carlo Algorithm for the Joint Inversion of Phase Velocity and Attenuation. (accepted to Geophysical Journal International).

Afanasiev M, Boehm C, van Driel M, Krischer L, Rietmann M, May DA, Knepley MG and Fichtner A (2019) Modular and flexible spectral-element waveform modelling in two and three dimensions. Geophysical Journal International, 216(3), 1675–1692. https://doi.org/10.1093/gji/ggy469.

Albarello D and Baliva F (2009) In-Situ Estimates of Material Damping from Environmental Noise Measurements. In Increasing Seismic Safety by Combining Engineering Technologies and Seismological Data (pp. 73–84). https://doi.org/10.1007/978-1-4020-9196-4_6

Anderson JG, Lee Y, Zeng Y and Day S (1996) Control of strong motion by the upper 30 meters. Bulletin of the Seismological Society of America, 86(6), 1749–1759. https://doi.org/10.1785/BSSA0860061749.

Aki K (1957). Space and Time Spectra of Stationary Stochastic Waves, with Special Reference to Microtremors. Bulletin of the Earthquake Research Institute, 35, 415–459.

Aki K and Richards PG (1980) Quantitative Seismology, Theory and Methods. Volume I: 557 pp., 169 illustrations. Volume II: 373 pp., 116 illustrations. San Francisco: Freeman. Price: Volume I, U.S. $35.00; Volume II, U.S. $35.00. ISBN 0 7167 1058 7 (Vol. I), 0 7167 1059 5 (Vol. II). Geological Magazine, 118(2), 208–208. https://doi.org/10.1017/S0016756800034439.

Badsar SA, Schevenels M, Haegeman W and Degrande G (2010) Determination of the material damping ratio in the soil from SASW tests using the half-power bandwidth method. Geophysical Journal International, 182(3), 1493–1508. https://doi.org/10.1111/j.1365-246X.2010.04690.x.

Badsar S (2012) In situ determination of material damping in the soil at small deformation ratios. Katholieke Universiteit Leuven.

Bergamo P, Marano S and Fah D (2023) Joint estimation of S-wave velocity and damping ratio of the near-surface from active Rayleigh wave surveys processed with a wavefield decomposition approach. Geophysical Journal International 233 (3), 1560–1579. https://doi.org/10.1093/gji/ggad010.

Biot MA (1956) Theory of Propagation of Elastic Waves in a Fluid-Saturated Porous Solid. I. Low-Frequency Range. The Journal of the Acoustical Society of America, 28(2), 168–178. https://doi.org/10.1121/1.1908239.

Capon J (1969) High-resolution frequency-wavenumber spectrum analysis. Proceedings of the IEEE, 57(8), 1408–1418. https://doi.org/10.1109/PROC.1969.7278.

Lai CG (1998) Simultaneous inversion of Rayleigh phase velocity and attenuation for near-surface site characterization. Georgia Institute of Technology.

Cupillard P and Capdeville Y (2010) On the amplitude of surface waves obtained by noise correlation and the capability to recover the attenuation: a numerical approach. Geophysical Journal International. https://doi.org/10.1111/j.1365-246X.2010.04586.x.

Comina C, Foti S, Boiero D and Socco LV (2011) Reliability of $V_S$, 30 evaluation from surface-wave tests. Journal of Geotechnical and Geoenvironmental engineering, 137(6), 579-586.

Cox BR and Beekman AN (2011) Intra-Method Variability in ReMi Dispersion and Vs Estimates at Shallow Bedrock Sites. Journal of Geotechnical and Geoenvironmental Engineering, 137(4), pp. 354-362.

Cox BR and Wood CM (2011) Surface Wave Benchmarking Exercise: Methodologies, Results, and Uncertainties. GeoRisk 2011, 845–852. https://doi.org/10.1061/41183(418)89.

Crow H, Hunter JA and Motazedian D (2011) Monofrequency in situ damping measurements in Ottawa area soft soils. Soil Dynamics and Earthquake Engineering, 31(12), 1669–1677. https://doi.org/10.1016/j.soildyn.2011.07.002.

Zywicki D (1999) Advanced signal processing methods applied to engineering analysis of seismic surface waves. Georgia Institute of Technology.

Johnson D and Dudgeon D (1993) Array signal processing: Concepts and techniques. Englewood Cliffs, NJ: P T R Prentice Hall.

Evans JP, McCalpin JP and Holmes DC (1996) Geologic Map of the Logan 7.5' Quadrangle, Cache County, Utah. Miscellaneous Publication 96-1, Utah Geological Survey, a division of Utah Department of Natural Resources, Salt Lake City, Utah. ISBN 1-55791-375-7.

Ewing WM, Jardetzky WS, Press F and Beiser A (1957) Elastic Waves in Layered Media. Physics Today, 10(12), 27–28. https://doi.org/10.1063/1.3060203.

Foti S (2004) Using transfer function for estimating dissipative properties of soils from surface-wave data. Near Surface Geophysics, 2(4), 231–240. https://doi.org/10.3997/1873-0604.2004020.

Foti S, Aimar M and Ciancimino A (2021) Uncertainties in Small-Strain Damping Ratio Evaluation and Their Influence on Seismic Ground Response Analyses (pp. 175–213). https://doi.org/10.1007/978-981-16-1468-2_9.

Foti S, Hollender F, Garofalo F, Albarello D, Asten M, Bard PY, Comina C, Cornou C, Cox B, Di Giulio G, Forbriger T, Hayashi K, Lunedei E, Martin A, Mercerat D, Ohrnberger M, Poggi V, Renalier F, Sicilia D and Socco V (2018) Guidelines for the good practice of surface wave analysis: a product of the InterPACIFIC project. Bulletin of Earthquake Engineering, 16(6), 2367–2420. https://doi.org/10.1007/s10518-017-0206-7.

Foti S, Lai C, Rix GJ and Strobbia C (2014) Surface Wave Methods for Near-Surface Site Characterization. CRC Press. https://doi.org/10.1201/b17268.

Haendel A, Ohrnberger M and Krüger F (2016) Extracting near-surface Q L between 1–4 Hz from higher-order noise correlations in the Euroseistest area, Greece. Geophysical Journal International, 207(2), 655–666. https://doi.org/10.1093/gji/ggw295.

Hall L and Bodare A (2000) Analyses of the cross-hole method for determining shear wave velocities and damping ratios. Soil Dynamics and Earthquake Engineering, 20(1–4), 167–175. https://doi.org/10.1016/S0267-7261(00)00048-8.

Herrmann RB (2013) Computer programs in seismology: an evolving tool for instruction and research, Seismol. Res. Lett., 84, 1081–1088. 10.1785/0220110096.

Holzlöhner U (1980) Vibrations of the elastic half-space due to vertical surface loads. Earthquake Engineering & Structural Dynamics, 8(5), 405–414. https://doi.org/10.1002/eqe.4290080504.

Johnston DH, Toksöz M N and Timur A (1979) Attenuation of seismic waves in dry and saturated rocks: II. Mechanisms. Geophysics, 44(4), 691–711. https://doi.org/10.1190/1.1440970

Jongmans D (1990) In-situ attenuation measurements in soils. Engineering Geology, 29(2), 99–118. https://doi.org/10.1016/0013-7952(90)90001-H.

Keilis-Borok V (1989) Seismic Surface Waves in a Laterally Inhomogeneous Earth (V. I. Keilis-Borok, Ed.; Vol. 9). Springer Netherlands. https://doi.org/10.1007/978-94-009-0883-3.

Lacoss RT, Kelly EJ and Toksöz MN (1969) Estimation of seismic noise structure using arrays. Geophysics, 34(1), 21–38. https://doi.org/10.1190/1.1439995

Lai CG, Rix GJ, Foti S and Roma V (2002) Simultaneous measurement and inversion of surface wave dispersion and attenuation curves. Soil Dynamics and Earthquake Engineering, 22(9–12), 923–930. https://doi.org/10.1016/S0267-7261(02)00116-1.

Lamb H (1904) On the propagation of tremors over the surface of an elastic solid. Proceedings of the Royal Society of London, 72(477–486), 128–130. https://doi.org/10.1098/rspl.1903.0029.

Aimar A (2022) Uncertainties in the estimation of the shear-wave velocity and the small-strain damping ratio from surface wave analysis. Politecnico di Torino.

Michaels P (1998) In Situ Determination of Soil Stiffness and Damping. Journal of Geotechnical and Geoenvironmental Engineering, 124(8), 709–719. https://doi.org/10.1061/(ASCE)1090-0241(1998)124:8(709).

Nazarian S, Stokoe KH and Hudson WR (1983) Use of spectral analysis of surface waves method for determination of moduli and thicknesses of pavement systems. Transportation Research Record, 38–45. https://api.semanticscholar.org/CorpusID:58935998.

O'doherty RF and Anstey NA (1971) Reflections on Amplitudes*. Geophysical Prospecting, 19(3), 430–458. https://doi.org/10.1111/j.1365-2478.1971.tb00610.x.

Ohrnberger M, Schissele E, Cornou C, Wathelet M, Savvaidis A, Scherbaum F, . . . Kind F (2004) Microtremor array measurements for site effect investigations: Comparison of analysis methods for field data crosschecked by simulated wavefields. 13th World Conference on Earthquake Engineering. Vancouver, B.C., Canada.

Papadopoulos M, François S, Degrande G and Lombaert G (2018) The influence of uncertain local subsoil conditions on the response of buildings to ground vibration. Journal of Sound and Vibration, 418, 200–220. https://doi.org/10.1016/j.jsv.2017.12.021.

Park CB, Miller RD and Xia J (1999) Multichannel analysis of surface waves. Geophysics, 64(3), 800–808. https://doi.org/10.1190/1.1444590.

Parolai S (2014) Shear wave quality factor Qs profiling using seismic noise data from microarrays. Journal of Seismology, 18(3), 695–704. https://doi.org/10.1007/s10950-014-9440-5.

Parolai S, Lai CG, Dreossi I, Ktenidou OJ and Yong A (2022) A review of near-surface QS estimation methods using active and passive sources. Journal of Seismology, 26(4), 823–862. https://doi.org/10.1007/s10950-021-10066-5.

Peterson JR (1993) Observations and modeling of seismic background noise. In. Albuquerque, New Mexico: United States Geological Survey.

Prieto GA, Lawrence JF and Beroza GC (2009) Anelastic Earth structure from the coherency of the ambient seismic field. Journal of Geophysical Research: Solid Earth, 114(B7). https://doi.org/10.1029/2008JB006067.

Perez M (1969) The Capability of the Utah State University Drainage Farm as an Irrigation and Drainage Demonstration Project. All Graduate Plan B and other Reports, Spring 1920 to Spring 2023. 623. https://doi.org/10.26076/6e25-d8ed.

Ricker N (1940) The form and nature of seismic waves and the structure of seismograms. Geophysics, 5(4), 348–366. https://doi.org/10.1190/1.1441816.

Rix GJ, Lai CG and Spang AW (2000) In Situ Measurement of Damping Ratio Using Surface Waves. Journal of Geotechnical and Geoenvironmental Engineering, 126(5), 472–480. https://doi.org/10.1061/(ASCE)1090-0241(2000)126:5(472).

Rix G, Lai C and Foti S (2001) Simultaneous Measurement of Surface Wave Dispersion and Attenuation Curves. Geotechnical Testing Journal, 24(4), 350. https://doi.org/10.1520/GTJ11132J.

Richart FE, Hall JR and Woods RD (1970) Vibrations of Soils and Foundations. Prentice-Hall.

Sanchez-Sesma FJ and Campillo MI (2006) Retrieval of the Green's Function from Cross Correlation: The Canonical Elastic Problem. Bulletin of the Seismological Society of America, 96(3), 1182–1191. https://doi.org/10.1785/0120050181.

Schevenels M, François S and Degrande G (2009) EDT: An ElastoDynamics Toolbox for MATLAB. Computers & Geosciences, 35(8), 1752–1754. https://doi.org/10.1016/j.cageo.2008.10.012.

Shibuya S, Mitachi T, Fukuda F and Degoshi T (1995) Strain Rate Effects on Shear Modulus and Damping of Normally Consolidated Clay. Geotechnical Testing Journal, 18(3), 365. https://doi.org/10.1520/GTJ11005J.

Snieder R, Wapenaar K and Wegler U (2007) Unified Green's function retrieval by cross-correlation; connection with energy principles. Physical Review E, 75(3), 036103. https://doi.org/10.1103/PhysRevE.75.036103.

Spencer TW, Edwards CM and Sonnad JR (1977) Seismic wave attenuation in nonresolvable cyclic stratification. Geophysics, 42(5), 939–949. https://doi.org/10.1190/1.1440773.

Stoll RD (1974) Acoustic Waves in Saturated Sediments. In Physics of Sound in Marine Sediments (pp. 19–39). Springer US. https://doi.org/10.1007/978-1-4684-0838-6_2.

Stokoe K, Rix G, Nazarian S (1989) In situ seismic testing with surface waves. 12th International Conference on Soil Mechanics and Foundation Engineering.

Tao Y and Rathje E (2019) Insights into Modeling Small-Strain Site Response Derived from Downhole Array Data. Journal of Geotechnical and Geoenvironmental Engineering, 145(7). https://doi.org/10.1061/(ASCE)GT.1943-5606.0002048.

Teague DP, Cox BR, Bradley B and Wotherspoon L (2018) Development of Deep Shear Wave Velocity Profiles with Estimates of Uncertainty in the Complex Inter-Bedded Geology of Christchurch, New Zealand. Earthquake Spectra, 34(2), 639-672. (https://doi.org/10.1193/041117EQS069M).

Teague DP, Cox BR, and Rathje ER (2018) Measured vs. Predicted Site Response at the Garner Valley Downhole Array Considering Shear Wave Velocity Uncertainty from Borehole and Surface Wave Methods. Soil Dynamics and Earthquake Engineering, 113(10), 339-355. https://doi.org/10.1016/j.soildyn.2018.05.031.

Tsai VC (2011) Understanding the amplitudes of noise correlation measurements. Journal of Geophysical Research, 116(B9), B09311. https://doi.org/10.1029/2011JB008483.

Tokimatsu K (1998) Geotechnical site characterization using surface waves. International Conference on Earthquake Geotechnical Engineering, ISBN 90-5410-581-X, 1333–1368.

Vantassel JP and Cox BR (2020) SWinvert: a workflow for performing rigorous 1-D surface wave inversions. Geophysical Journal International, 224(2), 1141–1156. https://doi.org/10.1093/gji/ggaa426.

Vantassel JP and Cox BR (2022) SWprocess: A Workflow for Developing Robust Estimates of Surface Wave Dispersion Uncertainty, Journal of Seismology, 26 (1): 731-756 (https://doi.org/10.1007/s10950-021-10035-y).

Vantassel JP (2021) jpvantassel/swprocess: v0.1.0b0. Zenodo https://doi.org/10.5281/zenodo.4584129.

Verachtert R, Lombaert G and Degrande G (2018) Multimodal determination of Rayleigh dispersion and attenuation curves using the circle fit method. Geophysical Journal International, 212(3), 2143–2158. https://doi.org/10.1093/gji/ggx523.

Walsh JB (1966) Seismic wave attenuation in rock due to friction. Journal of Geophysical Research, 71(10), 2591–2599. https://doi.org/10.1029/JZ071i010p02591.

Walsh JB (1968) Attenuation in partially melted material. Journal of Geophysical Research, 73(6), 2209–2216. https://doi.org/10.1029/JB073i006p02209.

Wathelet M, Jongmans D and Ohrnberger M (2004) Surface-wave inversion using a direct search algorithm and its application to ambient vibration measurements. Near surface geophysics, 2(4), 211-221.

Wathelet M, Guillier B, Roux P, Cornou C and Ohrnberger M (2018) Rayleigh wave three-component beamforming: signed ellipticity assessment from high-resolution frequency-wavenumber processing of ambient vibration arrays. Geophysical Journal International, 215(1), 507–523. https://doi.org/10.1093/gji/ggy286.

Wathelet M, Chatelain JL, Cornou C, Di Giulio G, Guillier B, Ohrnberger M and Savvaidis A (2020) Geopsy: A User-Friendly Open-Source Tool Set for Ambient Vibration Processing. Seismological Research Letters, 91(3), 1878--1889, doi: 10.1785/0220190360.

Williams JS (1962) Lake Bonneville: Geology of Southern Cache Valley, Utah. Geological Survey Professional Paper 257-C. United States Government Printing Office, Washington, D.C.

Xia J, Miller RD, Park CB and Tian G (2002) Determining Q of near-surface materials from Rayleigh waves. Journal of Applied Geophysics, 51(2–4), 121–129. https://doi.org/10.1016/S0926-9851(02)00228-8.

Zywicki DJ and Rix GJ (2005) Mitigation of Near-Field Effects for Seismic Surface Wave Velocity Estimation with Cylindrical Beamformers. Journal of Geotechnical and Geoenvironmental Engineering, 131(8), 970–977. https://doi.org/10.1061/(ASCE)1090-0241(2005)131:8(970).

Zywicki D (1999) Advanced signal processing methods applied to engineering analysis of seismic surface waves. Georgia Institute of Technology.